\documentclass[twocolumn]{aastex631}

\usepackage{natbib}
\usepackage{amsmath}
\usepackage{multirow}
\usepackage[normalem]{ulem}
\usepackage{xcolor}
\usepackage{xspace}
\usepackage{hyperref}
\usepackage{booktabs}
\usepackage{physics} 

\usepackage{enumitem}   
\usepackage{graphicx}
\usepackage{wrapfig}

\makeatletter
\def\@to{to}
\makeatother

\newcommand{\Porb}{P_{\mathrm{orb}}}
\newcommand{\Rp}{R_{\mathrm{p}}}
\newcommand{\Mstar}{M_*}
\newcommand{\SolarMass}{M_\odot}
\newcommand{\EarthRad}{R_\oplus}
\newcommand{\kepler}{\emph{Kepler}\xspace}
\newcommand{\gaia}{\emph{Gaia}\xspace}
\newcommand {\epos}  {\texttt{epos}}
\newcommand {\emcee}  {\texttt{emcee}}

\accepted{for publication in The Astronomical Journal; September 5th, 2022}

\shorttitle{Earths and super-Earths into the Habitable Zone} 
\shortauthors{Bergsten et al. 2022} 

\begin{document}

\title{The Demographics of \kepler's Earths and super-Earths into the Habitable Zone}

\correspondingauthor{Galen J. Bergsten}
\email{gbergsten@email.arizona.edu}

\author[0000-0003-4500-8850]{Galen J. Bergsten}
\affil{Lunar and Planetary Laboratory, The University of Arizona, Tucson, AZ 85721, USA}

\author[0000-0001-7962-1683]{Ilaria Pascucci}
\affil{Lunar and Planetary Laboratory, The University of Arizona, Tucson, AZ 85721, USA}

\author[0000-0002-1078-9493]{Gijs D. Mulders}
\affil{Facultad de Ingenier\'{i}a y Ciencias, Universidad Adolfo Ib\'{a}\~{n}ez, Av.\ Diagonal las Torres 2640, Pe\~{n}alol\'{e}n, Santiago, Chile}
\affil{Millennium Institute for Astrophysics, Chile}

\author[0000-0002-3853-7327]{Rachel B. Fernandes}
\affil{Lunar and Planetary Laboratory, The University of Arizona, Tucson, AZ 85721, USA}

\author[0000-0003-3071-8358]{Tommi T. Koskinen}
\affil{Lunar and Planetary Laboratory, The University of Arizona, Tucson, AZ 85721, USA}

\begin{abstract}
Understanding the occurrence of Earth-sized planets in the habitable zone of Sun-like stars is essential to the search for Earth analogues. Yet a lack of reliable \kepler detections for such planets has forced many estimates to be derived from the close-in ($2<\Porb<100$\,days) population, whose radii may have evolved differently under the effect of atmospheric mass loss mechanisms. In this work, we compute the intrinsic occurrence rates of close-in super-Earths ($\sim1-2\,\EarthRad$) and sub-Neptunes ($\sim2-3.5\,\EarthRad$) for FGK stars ($0.56-1.63\,\SolarMass$) as a function of orbital period and find evidence of two regimes: where super-Earths are more abundant at short orbital periods, and where sub-Neptunes are more abundant at longer orbital periods. We fit a parametric model in five equally populated stellar mass bins and find that the orbital period of transition between these two regimes scales with stellar mass, like $P_\mathrm{trans} \propto M_*^{1.7\pm0.2}$. These results suggest a population of former sub-Neptunes contaminating the population of Gyr-old close-in super-Earths, indicative of a population shaped by atmospheric loss. Using our model to constrain the long-period population of intrinsically rocky planets, we estimate an occurrence rate of $\Gamma_\oplus = 15^{+6}_{-4}\%$ for Earth-sized habitable zone planets, and predict that sub-Neptunes may be $\sim$\,twice as common as super-Earths in the habitable zone (when normalized over the natural log orbital period and radius range used). Finally, we discuss our results in the context of future missions searching for habitable zone planets. 
\end{abstract}

\keywords{Exoplanets (498) --- Exoplanet atmospheres (487) --- Habitable planets (695) --- Super Earths (1655) --- Mini Neptunes (1063)}


\section{Introduction} \label{sec:intro}

The modern field of exoplanet science is built in part upon the success of the \kepler mission \citep{Borucki2010, Borucki2017}, which detected thousands of extra-solar planets over the course of a $\sim$\,four-year duration. Both before and after the renaissance of exoplanet information provided by \kepler, many have explored questions about the formation, evolution, and occurrence of exoplanets.

One such open question is the occurrence rate (or frequency) of Earth-sized planets in the habitable zone of a Sun-like star, a quantity commonly denoted as $\eta_\oplus$. The habitable zone is conventionally defined as the orbital period range around a star for which a planet with a CO$_2$-H$_2$O-N$_2$ atmosphere could sustain liquid water on its surface \citep[e.g.,][]{Huang1959, Hart1978, Kopparapu2013}. Ideally, detections of reliable rocky planet candidates within a star's habitable zone would be used to directly estimate $\eta_\oplus$, where confirmed and candidate planets are weighted by the inverse of their completeness.

However, only a minute fraction of \kepler's small planet detections reside in the habitable zone \citep{Thompson2018}. This is likely not due to an absence of these planets, but rather due to the difficulties of detecting small planets at large orbital periods. As a transit mission, \kepler detections rely on the geometric probability that a planet will occult its host star, and this probability decreases with increasing orbital period. \kepler's completeness falls off drastically outside of $100$\,days, which means that very few terrestrial-sized habitable zone planet candidates were detected. Those that \kepler did find are typically of low reliability \citep[i.e., a high likelihood of being false alarms due to instrument statistics;][]{Thompson2018}, though reliability is difficult to estimate for year-long orbital period planets with only three transits during \kepler's short four-year duration. Nonetheless, many works attempt to use this sample directly, both with \citep[e.g.,][]{Bryson2021} and without \citep[e.g.,][]{Hsu2019} considerations of candidate reliability. The already-small sample size, poor mission completeness at long orbital periods, and variable treatment of reliability have led to an order-of-magnitude range in occurrence estimates with comparably large uncertainties (see e.g., the discussion in \citealp{Kunimoto&Matthews2020}). These discrepancies persist even when the occurrence is normalized over the planet radius and orbital period ranges used for the calculation (denoted as $\Gamma_\oplus$). 

A much larger sample of transiting exoplanets can be found at orbital periods far smaller than the habitable zone \citep{Thompson2018}. The majority of reliable small planet detections have orbital periods less than $100$\,days, well outside the $150$\,day innermost bound for the habitable zone of lower-mass K-type stars. To circumvent the low number of habitable zone detections, previous works have relied on fitting the period-radius distribution of the small close-in planet population, and extrapolating this to longer habitable zone orbital periods \citep[e.g.,][]{Petigura2013}. However, the close-in population of small planets does not come without caveats, as these planets may be shaped by processes which do not apply to the habitable zone. One such example process is the mechanism of planetary atmospheric mass loss.

Atmospheric loss entails that planets close-in to their host stars are exposed to large amounts of stellar radiation which can lead to a reduction of their atmospheric envelopes. One signature of this process is the ``hot-Neptune desert," or a dearth of Neptune-sized planet detections at very close-in ($< 4$\,days) orbital periods in the Gyr-old \kepler distribution \citep{Beauge2013}. Atmospheric loss is also likely to play a role in creating the observed ``radius valley" feature shown by \citet{Fulton2017} in the population of small close-in \kepler planets. The radius valley refers to a relative dearth of intermediate-sized planets ($\sim1.5-2\,\EarthRad$) separating an abundance of super-Earths $\left(\sim1-1.5\,\EarthRad\right)$ from an abundance of sub-Neptunes $\left(\sim2-3.5\,\EarthRad\right)$. Evidence for the radius valley has also been seen in \textit{K2} data \citep{KHU2020}, suggesting that this feature is caused by some physical process and not a quirk of the \kepler sample. These classifications are further motivated by density: in the subset of planets with both mass and radius estimates, those larger than $1.6\,\EarthRad$ are not likely to be rocky \citep{Rogers2015}.

Many works have explored the radius valley as an evolutionary feature carved out over time as intermediate-sized planets lose their atmospheres and appear observationally smaller in size \citep[e.g.,][]{Berger2020b, Sandoval2021}. In models of XUV photoevaporation \citep[e.g.,][]{Owen&Wu2013, Owen&Wu2017} and/or core-powered mass loss \citep[e.g.,][]{Ginzburg2016, Ginzburg2018, GuptaSchlichting2019,GuptaSchlichting2020}, the close-in population of super-Earths would be riddled with once-larger planets that lost their atmospheres, shrinking in planet radius. The distinction of ``close-in" is critical, because models of atmospheric loss are typically flux-limited \citep[see][]{Rogers2021} and thus have some decreasing orbital period dependence. This means that the most likely processes that sculpt the radius valley and hot-Neptune desert are not efficient at larger orbital periods $-$ most notably, the habitable zone. As a result, the radius distribution of close-in and habitable zone planets may evolve differently with orbital period. sub-Neptunes at large orbital periods will not evolve into super-Earths. Hence the occurrence of small close-in planets is not representative of small planets in the habitable zone.

\citet{Lopez&Rice2018} note that the use of separable and/or uncorrelated power laws in planet radius and orbital period for the population of close-in planets may lead to over-estimations of terrestrial planet occurrence at large orbital periods. Here, the planet radius distribution would be dominated by the close-in planets (many of which are remnant cores), while the orbital period distribution is dominated by sub-Neptunes. \citet{Pascucci2019} demonstrate and quantify this effect by excluding the subset of close-in planets that are most affected by atmospheric loss. Their Model \#6 removes planets with orbital periods less than $25$\,days (beyond which theoretical models predict minimal photoevaporation) and produces a decrease in $\eta_\oplus$ from $\sim40\%$ down to $\sim5\%$. Understanding how atmospheric loss shapes the size and orbital period distributions of Gyr-old planets is critical to understanding how the close-in population extends towards the habitable zone for estimating occurrence rates.

The atmospheric evolution of these planets is further coupled with differences in planet formation driving the overall orbital period distribution. Hence, characterizing how atmospheric loss shapes the planet size distribution, along with disentangling the processes of formation and evolution, is critical to understanding how the close-in population extends towards the habitable zone for estimating occurrence rates.

The aforementioned processes of planet formation and atmospheric evolution are not identical for every star-planet system, as there are many aspects which depend on stellar mass. In the context of planet formation, studies of young (Myr-old) stars show that protoplanetary disk mass reservoirs are stellar mass dependent \citep[e.g.,][]{Pascucci2016}. Additionally, exoplanet surveys of Gyr-old stars find occurrence rates for small planets which decrease for hotter (more massive) stars \citep{Howard2012, Mulders2015, Yang2020, He2021}. Host star mass is also relevant to atmospheric loss, as heavier/brighter stars provide more flux with which to heat up a planet and eventually remove part of its atmospheric envelope. Previous work by \citet{Wu2019} shows evidence for a radius valley which depends on stellar mass, where the boundary between super-Earths and sub-Neptunes moves to larger planet radii around heavier stars.

The convolution of the above factors makes constraining habitable zone occurrence a complicated endeavour, so it is not unsurprising that recent occurrence rate estimates span almost two orders of magnitude ($0.03 \lesssim \Gamma_\oplus \lesssim 1.4$; see Section~\ref{sec:GammaLit}). However, none of these works have attempted to study habitable zone occurrence as a function of stellar mass. Yet with the now-standard dataset of the final \kepler \texttt{DR25} release \citep{Thompson2018}, updated host star properties based on \textit{Gaia} data \citep{Berger2020b}, and a modernized treatment of reliability \citep{Bryson2020-OG-Reliability}, it is time for us to analyze the \kepler exoplanet population in the context of stellar mass dependence with insights from atmospheric loss.

One avenue of exploration is the interaction between the super-Earth and sub-Neptune populations as a function of orbital period. In this work, we confirm and quantify a transition between a short-period regime dominated by super-Earths and a longer-period regime dominated by sub-Neptunes, previously seen in both observations (see e.g., Fig. 7 in \citealp{Petigura2018}; Fig. 1 in \citealp{Pascucci2019}) and theory (see e.g., Fig. 2 in \citealp{GuptaSchlichting2019}; Fig. 7 in \citealp{RogersOwen2021}). We adopt a parametric approach to characterize the effects of atmospheric loss and the presence of remnant cores, which we explore in both the combined suite of FGK stars and as a function of stellar mass. Our analysis is the first to reveal a scaling between host star mass (hence luminosity) and the orbital period where this transition occurs.

In Section~(\ref{sec:epos}), we describe our sample and occurrence rate calculations, and introduce a metric of fractional occurrence to study the super-Earth versus sub-Neptune relationship. We detail our results in Section~(\ref{sec:results}) for a sample of stellar mass bins encompassing FGK stars, and explore how trends in occurrence rates differ with host star mass. In Section~(\ref{sec:discussion}), we discuss our findings in the context of atmospheric loss mechanisms, and calculate the occurrence of Earth-sized planets in the habitable zone for different bins of stellar mass. We conclude with a summary in Section~(\ref{sec:summary}), where we also consider ongoing and future efforts to refine our understanding of atmospheric loss and the search for Earth analogues in the habitable zones of Sun-like stars.


\section{Sample and Methods} \label{sec:epos}

Here we describe the methodology of our exoplanet occurrence rate study, and introduce the tools used to calculate occurrence rates with a dependence on stellar mass.

\subsection{Star and Planet Samples}\label{sec:samples}

Our stellar sample stems from the \citet{Berger2020a} \gaia-\kepler Stellar Properties Catalog, which uses photometry and \gaia parallaxes to derive stellar masses for 186,301 \kepler stars with a median uncertainty of 7\%. We limit our sample to contain only Sun-like or FGK type stars, encompassing a stellar mass range of $0.56 - 1.63\,\SolarMass$ based on the relation between spectral type and stellar mass for main-sequence stars in \citet{MamajekTable}. We exclude M dwarfs due to the small number of both M stars and confirmed planets orbiting M stars in the \kepler sample.

We limit our planet sample to include those with orbital periods of $2 \leq \Porb \leq 100$\,days, beyond which the completeness of small planets is low \citep[e.g.,][]{Fulton2017}. While theoretical models predict the effects of photoevapoaration to be negligible beyond 50\,days \citep{Owen&Wu2017}, we include planets out to 100\,days to allow for multiple data points at long orbital periods in log space. Because we wish to encompass the radius valley feature seen in the population of small planets within \kepler (e.g.,\citealp{Fulton2017}), we select planets with radii between $1 \leq \Rp \leq 3.5\,\EarthRad$. This choice is made to specifically include the definitions used in \citet{Fulton2017} to denote super-Earths ($1-1.75\,\EarthRad$) and sub-Neptunes ($1.75-3.5\,\EarthRad$). We exclude planets smaller than $1.0\,\EarthRad$, for which the \kepler completeness is low beyond even $30$\,days (see e.g., \citealp{Thompson2018, Mulders2018}). We also do not include planets beyond $3.5\,\EarthRad$, which is beyond the radius ``cliff" where there is a noticeable drop in planet occurrence. The process(es) that form the cliff are most likely related to the accretion phase from the gaseous disk (see e.g., the review by \citealp{Bean2021}) and are not relevant for smaller planets \citep{Kite2019}. 

Using the stellar masses presented in \citet{Berger2020a}, we also divide the FGK sample and corresponding planets into five stellar mass bins, with ranges chosen to have comparable numbers of stars per bin. Details of these bins are included in Table~(\ref{tab:MassBins}). The number of super-Earths or sub-Neptunes is listed in the last two columns and indicates all confirmed and candidate planets in a given stellar mass bin. Because bins were chosen based on stellar counts, the number of planet detections varies from bin to bin, but counts are still numerous enough for our statistical purposes. Additionally, \citet{Wu2019} has shown that the location of the valley in planet radius ($\Rp$) changes with stellar mass ($M_*$), and suggests a scaling where $R_\mathrm{p, adj.} = R_\mathrm{p} (\frac{\Mstar}{\SolarMass})^{1/4}$. We adopt this scaling to re-position the radius valley (and thus the boundary between super-Earths and sub-Neptunes) based on the average stellar mass of each bin. The normalized position is chosen to be $R_\mathrm{valley} = 2\,\EarthRad$ at $\Mstar = 1\,\SolarMass$ by visual inspection of the completeness-weighted two-dimensional planet orbital period and radius distribution - see Appendix~(\ref{sec:Appendix_RVscaling}). The $R_\mathrm{valley}$ values per stellar mass bin are provided in Table~(\ref{tab:MassBins}). Previous works (e.g., \citealp{VanEylen2018, Martinez2019}) have found evidence for a radius valley which scales inversely with orbital period, though the dependence is shallow (roughly $R_\mathrm{valley}\propto P^{-0.1}$) and not seen in our weighted period-radius distributions, so we do not consider such a scaling in this work.

\begin{deluxetable}{ccccc}
\tabletypesize{\footnotesize}
\tablewidth{20pt}
    \centering
    \tablecaption{Stellar Mass Bins and Planet Counts \label{tab:MassBins}}
    \tablehead{
    \colhead{$M_*$ Range} & \colhead{$N_*$} & \colhead{$R_\mathrm{valley}$} & \colhead{$N_\mathrm{sE}$}& \colhead{$N_\mathrm{sN}$} \\
     \colhead{$\left[M_\odot\right]$} &
      &
     \colhead{$\left[R_\oplus\right]$} 
     &
     &
    }
    \startdata{}
    0.56 - 0.81 & 22841 & 1.82 & 217 & 294 \\
    0.81 - 0.91 & 22929 & 1.93 & 186 & 230 \\
    0.91 - 1.01 & 22904 & 1.98 & 205 & 219 \\
    1.01 - 1.16 & 22912 & 2.04 & 237 & 197 \\
    1.16 - 1.63 & 23026 & 2.17 & 174 & 140 \\
    \enddata{}
    \tablecomments{We adopt a radius valley which scales with average stellar mass using the relation of \citet{Wu2019}, normalized to $R_\mathrm{valley} = 2\,\EarthRad$ at $M_* = 1.0\,\SolarMass$. The valley is considered the boundary between super-Earths and sub-Neptunes, with counts denoted under $N_\mathrm{sE}$ and $N_\mathrm{sN}$, respectively.}
\end{deluxetable}

\subsection{Exoplanet Occurrence and Fractional Occurrence}

To calculate exoplanet occurrence rates as a function of stellar mass, we use a modified version of \epos{}, the Exoplanet Population Observation Simulator \citep{Mulders2018}. \epos{} is a well documented and tested Python code developed by the ``Earths in Other Solar Systems" team\footnote{\url{http://eos-nexus.org/epos/}}, it is available on GitHub\footnote{\url{https://github.com/ GijsMulders/epos}}, and has already been used in several publications to compute occurrence rates as well as to compare planet formation models to the \kepler exoplanetary systems \citep[e.g.,][]{Kopparapu2018,Pascucci2018,Fernandes2019,Mulders2019,Pascucci2019,Mulders2020}.

In this work, we employ the classic inverse detection efficiency method (\texttt{Mode1} within \epos{}) for calculating planet occurrence rates. In short, a given planet's contribution to the occurrence of similar planets is weighted by a survey's ability to detect that specific planet, or the survey's ``completeness". For a survey with $N_*$ stars, the occurrence $\eta$ of planets in a bin of orbital period and radius is given by:
\begin{equation}
    \eta_\mathrm{bin} = \frac{1}{N_*} \sum_j^{N_\mathrm{pl}} \frac{1}{\mathrm{comp}_j}.
\end{equation}
Here, $\mathrm{comp}_j$ is the survey completeness evaluated at the orbital period and radius of the $j^\mathrm{th}$ planet in a bin containing $N_\mathrm{pl}$ planets. For details on the elements of survey completeness used in \kepler occurrence rates, see \citet{Mulders2018}. In a given bin, we treat the uncertainty on $\eta_\mathrm{bin}$ as the occurrence rate divided by the square root of the number of planet candidates in that bin.

Full details of the modifications to \epos{} for this work are described in Appendix~(\ref{epos4.0}), though we include a summary here for convenience. We have updated \epos{} to employ the \gaia-\kepler Stellar Properties Catalog \citep{Berger2020a}. Using these updated stellar properties when selecting a sample, \epos{} now accepts specified ranges of stellar mass and computes the relevant survey completeness maps internally - an external routine is also available\footnote{\url{https://github.com/gbergsten}}. In short, we adopt the methodology of \citet{Mulders2018} which includes generating 2D detection efficiency contours via \texttt{KeplerPORTS}\footnote{\url{https://github.com/nasa/KeplerPORTs}} \citep{BurkeCatanzarite2017}.

\epos{} uses the \kepler \texttt{DR25} catalog \citep{Thompson2018} to select confirmed and candidate planets as classified through an automated pipeline known as Robovetter \citep{Coughlin2017}, which also provides disposition scores. These scores describe the fraction of iterations in which a threshold crossing event is classified as a planet candidate, with iterations based on perturbing the event's Robovetter metrics about their uncertainties (see \citealp{Thompson2018} for further details). Disposition score cuts have been used to produce more reliable samples of planet candidates, though recent work by \citet{Bryson2020} suggests that score cuts are an imperfect metric which loses diagnostic value upon the implementation of a direct reliability treatment. 

As such, \epos{} now includes an optional treatment of vetting reliability as described by \citet{Bryson2020}. We calculate each candidate's reliability score as the product of $R_\mathrm{FA} \cdot (1-\mathrm{FPP})$, where $R_\mathrm{FA}$ is the false alarm reliability calculated for this work following the methods of \citealp{Bryson2020}, and $\mathrm{FPP}$ is the false positive probability calculated by \citet{Morton2016}. We then take each candidate's reliability score and treat this as a multiplicative factor on the occurrence calculated through the inverse detection efficiency method.

In addition to calculating the occurrence rates of exoplanets, we introduce a metric of ``fractional occurrence" to better describe the relative behavior between the super-Earth and sub-Neptune populations. Here, either the super-Earth or sub-Neptune occurrence distributions are divided by the combined occurrence of both groups. This can be thought of as the relative spacing between the two group's occurrence profiles with orbital period, where the \textit{fractional} occurrence of super-Earths defines how much they contribute to the \textit{combined} occurrence of both super-Earths and sub-Neptunes. Taking a ratio allows us to divide out any broader distributions with orbital period affecting both groups. Such an estimation can be affected by detection limits (larger planets are more easily found than smaller planets at larger orbital periods), but we will next employ a forward model that is less affected by this bias in order to show a robust feature of the planet population.


\subsection{Parametric Modeling of Exoplanet Distributions} \label{sec:model}

The inverse detection efficiency method described in the previous section is intuitive and computationally efficient, but not without shortcomings: it cannot predict occurrence rates in regimes without planet detections, and results may still harbor observational biases in areas of poor completeness. To complement this, we also calculate occurrence rates via a forward model using a parametric distribution to describe our planet population. This approach is more statistically sound, at the cost of increased computational time and specific assumptions on the functional form of the underlying planet distribution.

Previous works have shown that features of the \kepler planet population are well-described by (broken) power laws when using parametric distributions. In this work, we adopt the approach of \citet{Youdin2011} and \citet{Burke2015}. This method is chosen for their Poisson likelihood approach (useful for detections of varying sensitivity), and implementation of Bayesian Markov chain Monte Carlo (MCMC) parameter estimation (useful for fitting more complex parametric models). We briefly summarize here the salient components and refer the reader to Appendix~(\ref{FitDetails}) for details. 

Following \citet{Youdin2011}, we adopt a parametric description of the planet distribution function (PLDF) of the form:
\begin{equation} \label{eqn:df}
    \frac{\mathrm{d}^2 f}{\mathrm{d} \Porb \mathrm{d} \Rp} = F_\mathrm{0} C_\mathrm{n} g(\Porb,\Rp).
\end{equation}
where $F_\mathrm{0}$ can be treated as the average number of planets per star, $C_\mathrm{n}$ is a normalization factor, and $g(\textbf{x})$ is the highly customizable shape function used to describe the behavior of the planet distribution.

In our case, the shape function consists of two components, $g(\textbf{x}) = g_1 \cdot g_2$. The first is a broken power law in orbital period which governs the overall distribution of the combined super-Earth and sub-Neptune population:
\begin{equation}\label{eqn:bpl}
    g_1(\Porb) = 
    \begin{cases}
    (\Porb / P_\mathrm{break})^{\beta_1} & \text{if $P < P_\mathrm{break}$} \\
    (\Porb / P_\mathrm{break})^{\beta_2} & \text{if $P \geq P_\mathrm{break}$.}
    \end{cases}
\end{equation}
The exponents $\beta_1$ and $\beta_2$ govern the slope of the power law on either side of a break in orbital period $P_\mathrm{break}$. This form has been shown to reproduce well the overall occurrence of small Kepler planets vs. orbital period \citep[e.g.,][]{Petigura2018, Mulders2018}. 

The second component governs the fractional occurrence. We will first describe the functional form of this component here, then later motivate our choice of parameterization with \kepler occurrence rates in Section~(\ref{sec:observed_turnover}) and fits to the planetary parameters in Section~(\ref{sec:fitResults}). We choose to model the fractional occurrence of super-Earths via a functional form which simultaneously constrains three regimes: short orbital periods, long orbital periods, and a transition between the two. For this, we adopt the form of a hyperbolic tangent (similar to an ``S-curve") in $\log_{10} \Porb$, which allows for the transition between two asymptotic values via a smooth curve centered on a specific orbital period. The transition function itself includes two free parameters - one for the location of the central period of this curvature $P_\mathrm{central}$, and one for the log-period width $s$ between the central period and an asymptotic plateau on either side - and is given by the form: 
\begin{equation}
    t(\Porb) = 0.5 - 0.5 \tanh{\left(\frac{\log_{10}{\Porb} - \log_{10}{P_\mathrm{central}}}{\log_{10}{s}}\right)}.
\end{equation}

The plateaus both before and after the transition can be normalized independently, for which we adopt two additional free parameters. The fraction $\chi_1$ represents the asymptotic fractional occurrence of super-Earths at short orbital periods (well before $P_\mathrm{central}$), and the fraction $\chi_2$ denotes a similar asymptote at long orbital periods (well after $P_\mathrm{central}$). In this sense, the smoothness parameter $s$ governs how quickly the curve levels off towards $\chi_1$ or $\chi_2$ when moving away from $P_\mathrm{central}$. By definition, fractional occurrence compares either super-Earths or sub-Neptunes to the combined occurrence of both, such that the two size regimes are always complementary (i.e., sum to unity). Thus, at short periods sub-Neptunes should approach $\left(1-\chi_1\right)$, and at long periods approach $\left(1-\chi_2\right)$.

\begin{figure}
    \centering
    \includegraphics[width=0.42\textwidth]{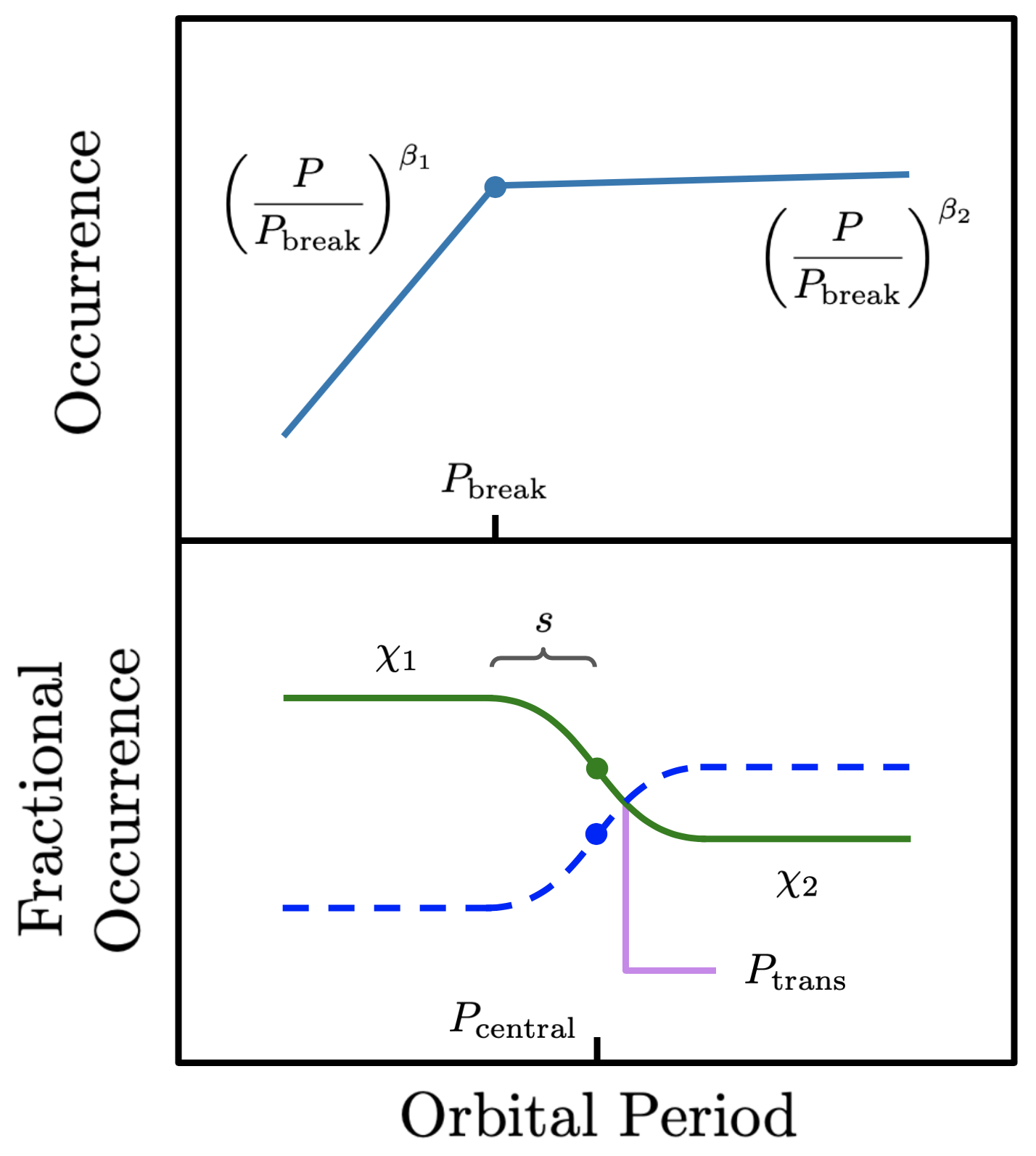}
    \caption{Graphical representation of the functional form adopted in this work to describe the close-in population of exoplanets. \textbf{Top:} A broken power law governs the overall orbital period distribution. \textbf{Bottom:} A transition function describes how the fractional occurrence behaves with period for super-Earths (green; $1\,\EarthRad - R_{valley}$) and sub-Neptunes (blue; $R_{valley} - 3.5\,\EarthRad$). Dots represent the relevant period free parameters (\textbf{top:} $P_\mathrm{break}$; \textbf{bottom:} $P_\mathrm{central}$). The purple line indicates the point $P_\mathrm{trans}$ where sub-Neptunes become more common than super-Earths.}
    \label{fig:sketch}
\end{figure}

Using the function $t = t(\Porb)$ to transition between short- and long-period regimes, we define the component governing fractional occurrence via a piece-wise function:
\begin{equation}
    G(\Porb, \Rp) = \begin{cases}
    t \cdot \chi_1 + \left[1 - t\right] \cdot \chi_2 & \text{if $\Rp < R_\mathrm{valley}$} \\
    \mkern-0mu\begin{array}{@{}l@{}}t \cdot \left(1-\chi_1\right) + \\ \quad{}\left[1 - t\right] \cdot \left(1-\chi_2\right)\end{array} & \text{if $\Rp > R_\mathrm{valley}$} \\
    \end{cases}
\end{equation}
which is represented graphically in Figure~(\ref{fig:sketch}) alongside the broken power law component $g_1$. Implementing the behavior of $G$ into the shape function via $g_2$ is not trivial, as constraining fractional occurrence requires the calculation of occurrence for both super-Earths and sub-Neptunes, which is the result of the PLDF itself - details of this implementation are included in Appendix~(\ref{FitDetails}). We opt to consider and fit this fractional occurrence metric, rather than fitting super-Earth and/or sub-Neptune occurrence rates individually, to reduce the number of free parameters.

To summarize, the eight free parameters of our model are:
\begin{itemize}
    \item $F_\mathrm{0}$: Average number of planets per star.
    \item $P_\mathrm{break}$: Location of the break in the orbital period power law.
    \item $\beta_1$: Exponent governing the power law for orbital periods $P < P_\mathrm{break}$.
    \item $\beta_2$: Exponent governing the power law for orbital periods $P \geq P_\mathrm{break}$
    \item $P_\mathrm{central}$: Central point of the hyperbolic tangent transition curve.
    \item $s$: Smoothness of the transition function about $P_\mathrm{central}$.
    \item $\chi_1$: Fractional occurrence of super-Earths at $P \ll P_\mathrm{central}$; in the same regime, sub-Neptune fractional occurrence is given by the complement $\left(1-\chi_1 \right)$.
    \item $\chi_2$: Fractional occurrence of super-Earths at $P \gg P_\mathrm{central}$; in the same regime, sub-Neptune fractional occurrence is given by the complement $\left(1-\chi_2 \right)$.
\end{itemize}
The PLDF is fit to the \kepler \texttt{DR25} sample through maximum likelihood estimation techniques in an MCMC approach using \emcee{} \citep{ForemanMackey2012}. While we do not expressly fit a dependence in stellar mass using a PLDF, we utilize the methods of \citet{Youdin2011} and \citet{Burke2015} to describe the planet population in each stellar mass bin independently. As such, we stress that the value and interpretation of each parameter is dependent on the relevant mass range (e.g., $F_0$ in the lightest stellar mass bin should be considered the average number of planets per star \textit{for stars with $0.58 < M_* < 0.81\,\SolarMass$}).

As noted in \citet{Bryson2020-OG-Reliability}, the methodology of \citet{Youdin2011} and \citet{Burke2015} considers survey completeness but not reliability. To implement the latter, we follow the approach of \citet{Bryson2020-OG-Reliability} by performing multiple independent fits where the input population of planets is drawn based on their reliability (e.g., a planet candidate with a reliability of $0.8$ is included in $80\%$ of these fits). In each mass bin, we perform $100$ fits, each with $16$ walkers run for $10,000$ steps (discarding $500$ steps of burn-in), then concatenate the posteriors to produce reliability-informed free parameter distributions. The impacts of our reliability considerations are discussed in Appendix~(\ref{ReliabilityEffects}).


\section{Results} \label{sec:results}

Here we report the results of our occurrence rate calculations, first using the inverse detection efficiency method, followed separately by the results from our parametric modeling.

\subsection{Occurrence Rates from the Inverse Detection Efficiency Method} \label{sec:occResults}

\begin{figure}
    \centering
    \includegraphics[width=0.425\textwidth]{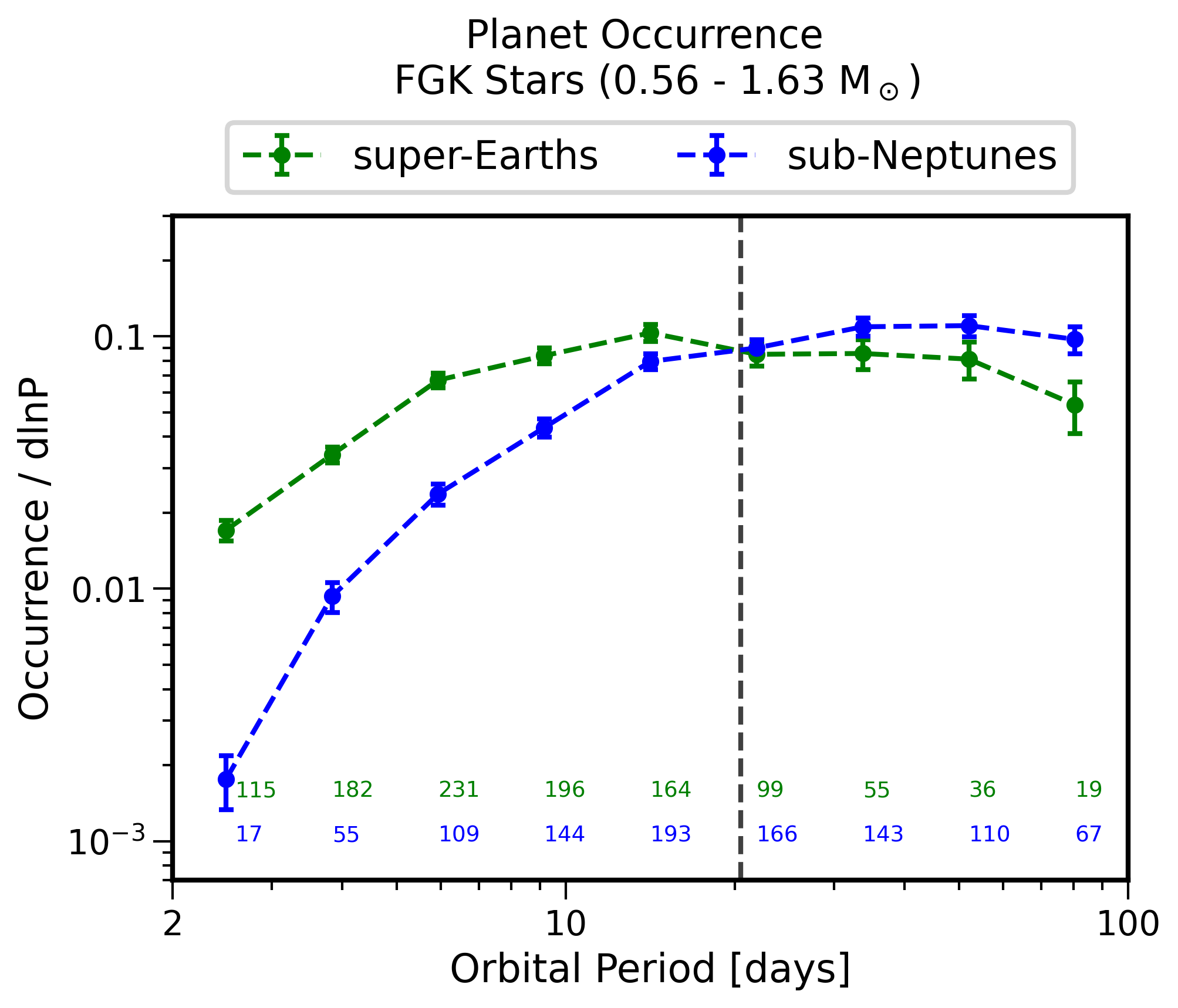}
    \caption{The occurrence of super-Earths (green; $1\,\EarthRad - R_{valley}$) and sub-Neptunes (blue; $R_{valley} - 3.5\,\EarthRad$) as a function of orbital period, calculated for all planets in the full FGK sample ($0.56 - 1.63\,\SolarMass$). The colored numbers indicate the number of observed candidate planets of a given type within a particular orbital period bin. The black vertical line indicates the approximate location where the super-Earth and sub-Neptune curves cross.}
    \label{fig:EPOS_OccANDFracOcc}
\end{figure}

We calculate the occurrence of super-Earths ($1-2\,\EarthRad$) and sub-Neptunes ($2-3.5\,\EarthRad$) separately in several orbital period bins for our entire FGK sample with \epos{}. When comparing the occurrence of super-Earths and sub-Neptunes as a function of orbital period (Fig.~\ref{fig:EPOS_OccANDFracOcc}), we note that at short orbital periods ($\Porb < 20$\,days) super-Earths are more common than sub-Neptunes, while at longer orbital periods ($\Porb > 20$\,days) the converse is true.

This trend has been shown previously by \citet{Petigura2018} who characterize hot versus warm planets with stellar metallicity, and by \citet{Pascucci2019} while investigating the impact of bare cores in $\eta_\oplus$ estimates. The aforementioned works observe this transition at an orbital period of $10$\,days while using a radius valley of $1.8\,\EarthRad$. Our increased estimate of $20$\,days stems from placing the radius valley at $\sim2.0\,\EarthRad$ in Figure~(\ref{fig:EPOS_OccANDFracOcc}) due to our adoption of a radius valley which scales with average stellar mass. The qualitative result is the same: there is an orbital period-dependence and transition which separates the super-Earth and sub-Neptune dominated regimes.

\subsubsection{A stellar-mass dependent transition period from super-Earths to sub-Neptunes} \label{sec:observed_turnover}

\begin{figure}
    \centering
    \includegraphics[width=0.425\textwidth]{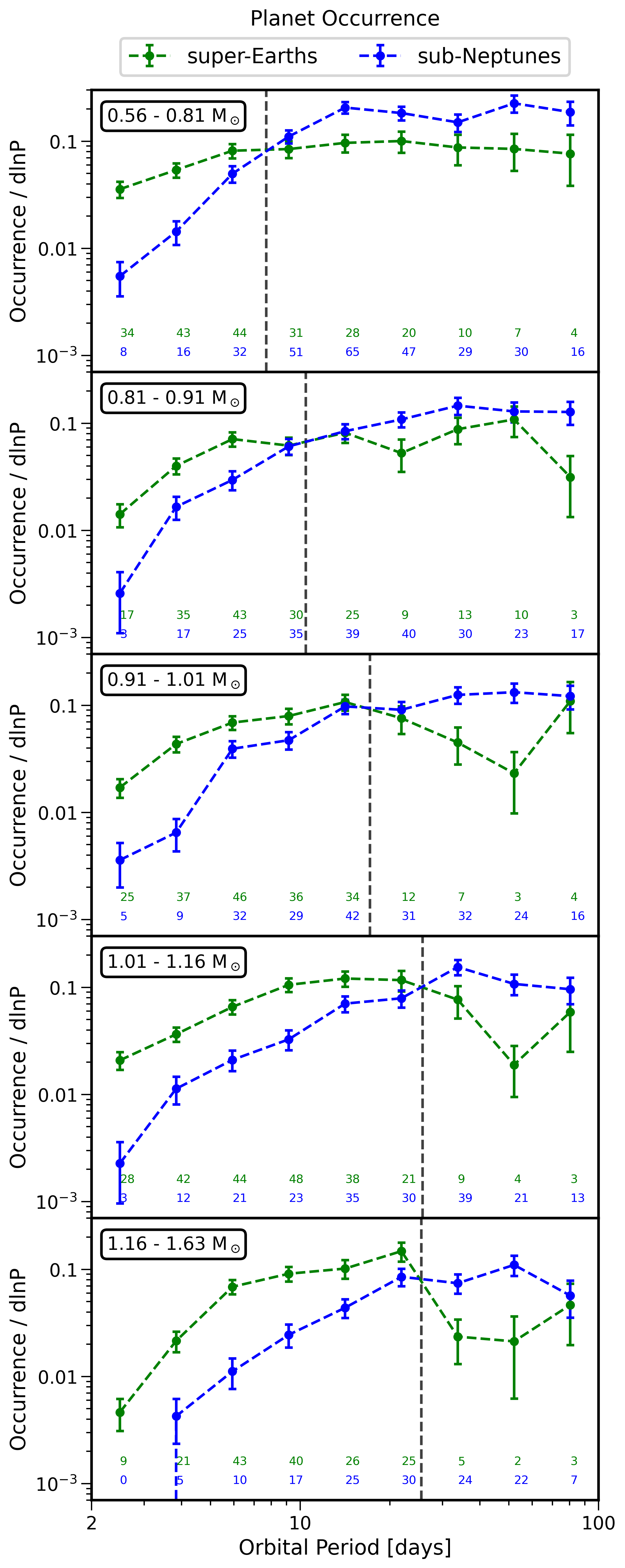}
    \caption{The occurrence of super-Earths (green; $1\,\EarthRad - R_{valley}$) and sub-Neptunes (blue; $R_{valley} - 3.5\,\EarthRad$) as a function of orbital period. Each panel represents a different stellar mass bin, increasing from top to bottom. The colored numbers in each panel indicate the number of observed candidate planets of a given type within a particular orbital period bin (and stellar mass bin). Black vertical lines indicate the approximate orbital period where the occurrence curves cross.}
    \label{fig:EPOS_Occ}
\end{figure}

We then test if these two distinct orbital period regimes persist as a function of stellar mass by repeating our occurrence rate calculations for each of our five stellar mass bins. The occurrence of super-Earths and sub-Neptunes vs. orbital period per stellar-mass bin is presented in Figure~(\ref{fig:EPOS_Occ}), and the details on bin ranges, counts, and planet samples can be found in Table~(\ref{tab:MassBins}). The same trend seen in the FGK sample can be seen in each stellar mass bin, suggesting it is a common feature of small planets around dwarf stars. We note that the orbital period out to which super-Earths are more abundant shows a stellar mass dependence: sub-Neptunes become more common than super-Earths within $10$\,days for the lowest mass bins (where $R_\mathrm{valley}\approx1.8\,\EarthRad$, in agreement with \citealp{Petigura2018} and \citealp{Pascucci2019}), but not until $\Porb \approx 20$\,days for the heavier mass bins. This may be indicative of an atmospheric loss front (or ``cosmic shoreline," \citealp{ZahnleCatling2017}) which produces more remnant cores out to larger orbital periods around heavier, more luminous stars.

\begin{figure*}
    \centering
    \includegraphics[scale=0.4]{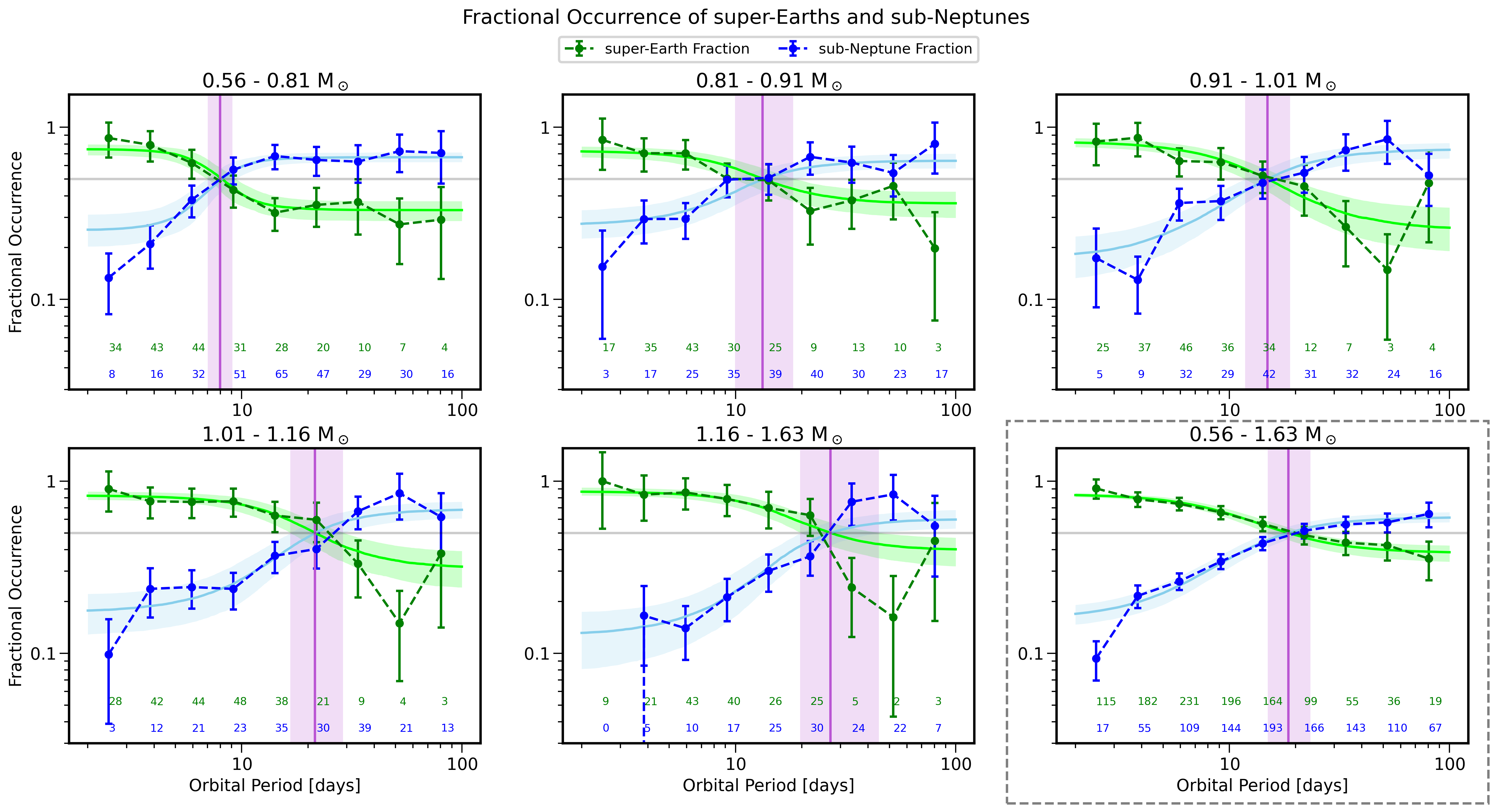}
    \caption{The fractional occurrence of super-Earths (green) and sub-Neptunes (blue) relative to their combined planet occurrence. Each panel represents a different stellar mass bin, with the full FGK sample in the lower-right. The colored numbers in each panel indicate the number of observed candidate planets of a given type within a particular orbital period bin (and stellar mass bin). Dashed points represent the observed occurrence as calculated by \epos{} for a selection of orbital period bins. Solid lines indicate the best fit distributions as evaluated for that bin. The horizontal grey line marks a fractional occurrence of $0.5$, corresponding to an equal abundance of super-Earths and sub-Neptunes, and the orbital period where each fit matches this condition is marked by a vertical purple line. Shaded regions indicate $1\sigma$ envelopes.}
    \label{fig:FracOcc}
\end{figure*}

To better understand this variation in orbital period, we employ our metric of fractional occurrence. The super-Earth or sub-Neptune occurrence rates from \epos{} are divided by the sum of both, and are presented in Figure~(\ref{fig:FracOcc}) for both the entire FGK sample and each stellar mass bin. Again two distinct regimes can be seen, where the short-period super-Earth dominated regime and long-period sub-Neptune dominated regime are separated by a transition period which increases with stellar mass.

A small number of planets detected at large orbital periods leads to an apparent upturn with wide uncertainties in the fractional occurrence of super-Earths (or downturn in the sub-Neptunes) for bins with $M_* > 0.91\,\SolarMass$ (see Figure~\ref{fig:FracOcc}). The upturn is not present in the \epos{} calculation of fractional occurrence for the combined FGK sample (lower right panel of Figure~\ref{fig:FracOcc}), so we do not comment with certainty on the implications of this trend.

\subsection{Occurrence Rates from the Parametric Forward Model} \label{sec:fitResults}

\renewcommand{\arraystretch}{1.5}
\begin{deluxetable*}{cccccccccccc}
\tabletypesize{\footnotesize}
\tablewidth{0pt}
    \tablecaption{Parameter Fits with $68\%$ Uncertainties for the Distribution Function from Section~(\ref{sec:model}) \label{tab:fits}}
    \tablehead{
    \colhead{$M_*$ Range} & \colhead{$F_\mathrm{0}$} & \colhead{$P_\mathrm{break}$} & \colhead{$\beta_1$}& \colhead{$\beta_2$} & \colhead{$P_\mathrm{central}$} &\colhead{$s$} &\colhead{$\chi_1$} &\colhead{$\chi_2$} \\
    $\left[M_\odot\right]$ & & [days] & & & [days] & [days] & & 
    }
    \startdata{}
    0.56 - 0.81 & $0.89_{-0.05}^{+0.05}$ & $14.29_{-2.76}^{+2.63}$ & $0.15_{-0.11}^{+0.15}$ & $-1.15_{-0.17}^{+0.17}$ & $7.54_{-0.84}^{+0.76}$ & $1.46_{-0.31}^{+0.28}$ & $0.75_{-0.05}^{+0.05}$ & $0.33_{-0.04}^{+0.04}$ \\
    0.81 - 0.91 & $0.70_{-0.05}^{+0.05}$ & $6.13_{-1.37}^{+2.57}$ & $1.19_{-0.54}^{+0.83}$ & $-0.68_{-0.13}^{+0.10}$ & $11.26_{-2.35}^{+2.87}$ & $2.02_{-0.47}^{+0.62}$ & $0.73_{-0.05}^{+0.05}$ & $0.36_{-0.06}^{+0.06}$ \\
    0.91 - 1.01 & $0.63_{-0.04}^{+0.04}$ & $6.92_{-0.85}^{+1.25}$ & $0.91_{-0.31}^{+0.32}$ & $-0.84_{-0.11}^{+0.09}$ & $12.98_{-2.34}^{+2.22}$ & $2.53_{-0.48}^{+0.48}$ & $0.83_{-0.05}^{+0.05}$ & $0.26_{-0.08}^{+0.08}$ \\
    1.01 - 1.16 & $0.61_{-0.04}^{+0.04}$ & $12.02_{-1.34}^{+1.78}$ & $0.44_{-0.14}^{+0.15}$ & $-1.10_{-0.15}^{+0.13}$ & $17.57_{-3.21}^{+3.59}$ & $2.16_{-0.48}^{+0.57}$ & $0.83_{-0.05}^{+0.05}$ & $0.31_{-0.08}^{+0.08}$ \\
    1.16 - 1.63 & $0.50_{-0.04}^{+0.04}$ & $6.96_{-0.82}^{+0.70}$ & $1.90_{-0.39}^{+0.50}$ & $-0.69_{-0.11}^{+0.10}$ & $16.57_{-2.83}^{+3.62}$ & $2.15_{-0.44}^{+0.54}$ & $0.87_{-0.05}^{+0.05}$ & $0.39_{-0.08}^{+0.07}$ \\
    \hline
    0.56 - 1.63 & $0.63_{-0.02}^{+0.02}$ & $9.98_{-1.79}^{+1.42}$ & $0.58_{-0.12}^{+0.20}$ & $-0.90_{-0.09}^{+0.09}$ & $11.33_{-1.36}^{+1.71}$ & $2.49_{-0.34}^{+0.41}$ & $0.84_{-0.02}^{+0.02}$ & $0.38_{-0.04}^{+0.04}$ 
    \enddata{}
    \tablecomments{Each stellar mass bin is fit individually for the average number of planets per star $F_0$, exponents $\beta_1$ and $\beta_2$ in a broken power law on either side of $P_\mathrm{break}$, and smoothness $s$ of a transition from $\chi_1$ to $\chi_2$ about $P_\mathrm{central}$.}
\end{deluxetable*}

The PLDF in Equation~(\ref{eqn:df}) is fit to the planet sample via MCMC independently for each stellar mass bin, and we present the median and $1\sigma$ posteriors of each parameter in Table~(\ref{tab:fits}). We use the distributions of these best-fit parameters to evaluate Equation~(\ref{eqn:df}) over the radius ranges corresponding to super-Earths and sub-Neptunes to calculate the posterior distributions of their relative fractional occurrence as a function of orbital period. We plot the median and $1\sigma$ curves from these distributions in Figure~(\ref{fig:FracOcc}), where we show how our forward model for calculating occurrence rates compares to the fractional occurrence trends seen with the inverse detection efficiency method. We find that our choice to account for reliability improves consistency between the two methods in some bins while widening discrepancies in others --- further discussion and an additional comparison \textit{without} reliability are included in Appendix~(\ref{ReliabilityEffects}). In both cases, the two methods are consistent (at the $1\sigma$ level) for the majority of bins, so we conclude that our adopted shape function provides a reasonable match to observed trends in fractional occurrence.

In Appendix~(\ref{sec:pldf}), we integrate Equation~(\ref{eqn:df}) over the radius range of our sample to affirm that our model is able to capture known trends in the overall occurrence distribution of small planets (in addition to new trends in fractional occurrence) seen with the inverse detection efficiency method. We find that the adopted broken power law form of $g_1$ reproduces well the overall small planet distribution with orbital period and comment on key values in the Appendix.

\subsubsection{Parametric Dependence on Stellar Mass}\label{sec:ParameterDetails}

Because the PLDF approach allows us to characterize the small planet population and corresponding occurrence trends independently for each stellar mass bin, we are able to compare the best-fit parameters as a function of stellar mass. The results of our fitting suggest three parameters with stellar mass dependencies, which we describe here.

First, we find that the average number of planets per star, $F_0$, decreases with increasing stellar mass (see 2nd column in Table~\ref{tab:fits}). Previous works \citep{Howard2012, Mulders2015, Yang2020, He2021} have found similar trends for small planets within the FGK sample using stellar effective temperature, which is a good proxy for stellar mass when dealing with main-sequence stars. Each work considers different ranges of planet orbital period and radius space, but produces the same qualitative trend seen in this work: an anti-correlation between the number of small planets per star versus stellar temperature and thus stellar mass.

\begin{figure}
    \centering
    \includegraphics[width=0.45\textwidth]{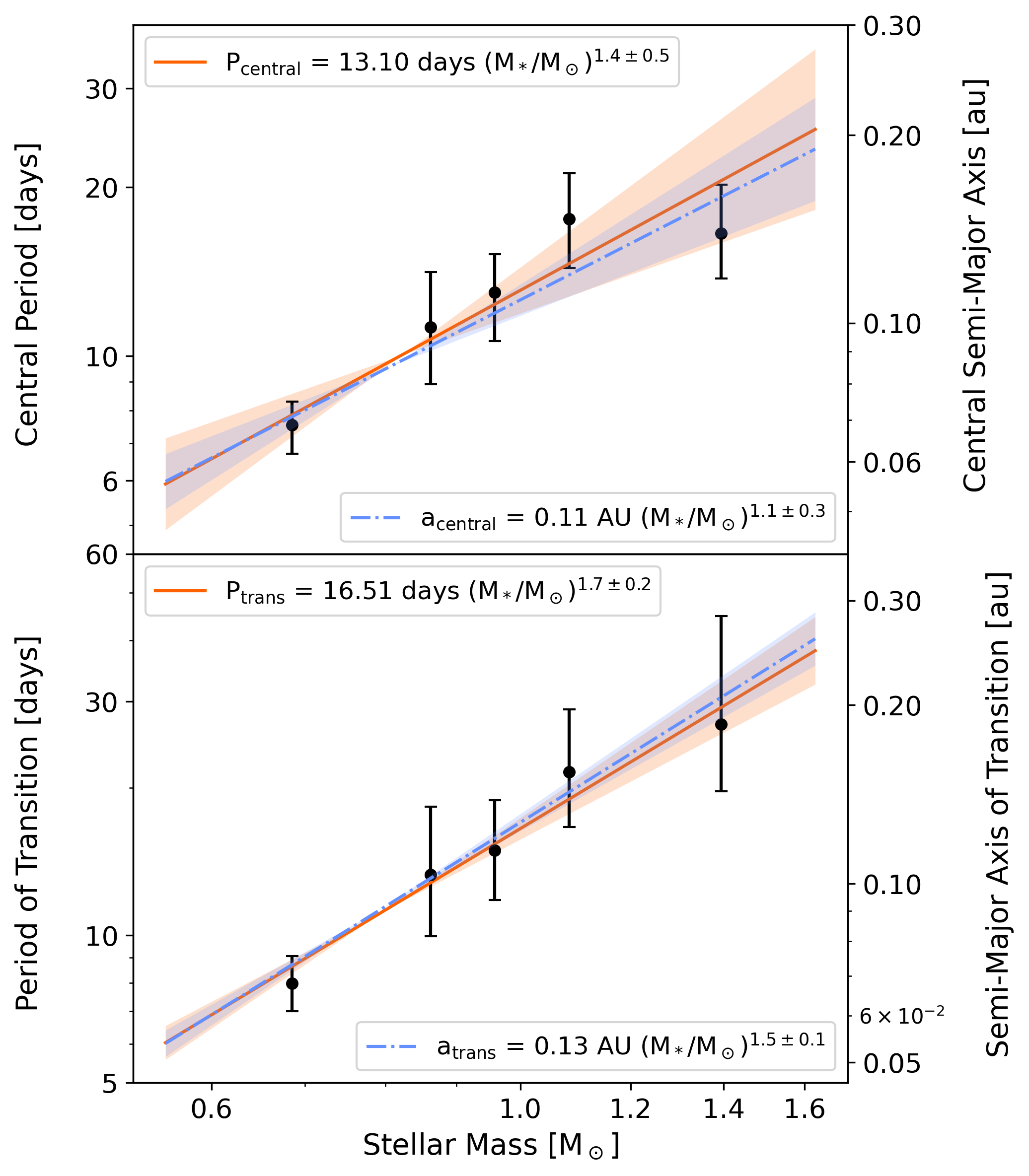}
    \caption{\textbf{Top:} The best-fit values of $P_\mathrm{central}$ as calculated for each stellar mass bin. Vertical error bars represent the $1\sigma$ uncertainties in $P_\mathrm{central}$ as drawn from their posteriors. In orange, we present log-log fits between stellar mass and orbital period, which has a dependence on stellar mass of $\alpha = 1.4 \pm 0.5$. \textbf{Bottom:} Similar to above, but now plotting the median and $1\sigma$ uncertainties of $P_\mathrm{trans}$ evaluated from the best-fit fractional occurrence curves; the orange line represents represents a log-log fit between stellar mass and transition period with $\alpha = 1.7\pm0.2$. In both panels, the rightmost vertical axis presents the same information converted to semi-major axis, with log-log fits in dot-dashed blue (\textbf{top:} $\alpha = 1.1 \pm 0.3$; \textbf{bottom:} $\alpha = 1.5 \pm 0.1$).}
    \label{fig:Mass-vs-Turnover}
\end{figure}

Second, we note that the orbital period $P_\mathrm{central}$ (i.e. the center of curvature separating the short- and longer-period regimes) depends on stellar mass. There is a clear increase from $P_\mathrm{central} = 8_{-1}^{+1}$\,days at the lowest stellar mass bin $(0.56-0.82\,\SolarMass)$ to $P_\mathrm{central} \approx 17_{-3}^{+4}$\,days in the heaviest stellar mass bin $(1.16-1.63\,\SolarMass)$. We fit a log-log relationship to the best-fit values reported in Table~(\ref{tab:fits}) and find that $P_\mathrm{central} \propto \Mstar^{1.4 \pm 0.5}$, shown in Figure (\ref{fig:Mass-vs-Turnover}). We also convert the best-fit central periods to semi-major axes using \kepler's third law, employing the average stellar mass for each bin. A log-log relationship is again fit to return a modified exponent of $a_\mathrm{central} \propto \Mstar^{1.1 \pm 0.3}$, producing as expected a slightly shallower relationship with semi-major axis $a$.

However, while the central period is insightful to a shift in behavior within the super-Earth and sub-Neptune populations, the precise period location is contingent on our adopted parameterization (i.e., a hyperbolic tangent). A more robust metric could be the orbital period where super-Earths and sub-Neptunes are present in equal abundance, which also marks the transition where the dominant species switches from the former to the latter. In each stellar mass bin, we evaluate the distribution of this transition period (hereafter $P_\mathrm{trans}$) from the posterior fractional occurrence distributions, plotting the median and $1\sigma$ ranges in Figure~(\ref{fig:FracOcc}). To estimate how $P_\mathrm{trans}$ scales with stellar mass, we repeat the above procedure for fitting log-log relationships and plot the results in the lower panel of Figure (\ref{fig:Mass-vs-Turnover}). We highlight a steeper stellar mass dependence as compared to $P_\mathrm{central}$, noting that $P_\mathrm{trans} \propto \Mstar^{1.7 \pm 0.2}$, which persists in semi-major axis as $a_\mathrm{trans} \propto \Mstar^{1.5 \pm 0.1}$.

Finally, we note that the fraction $\chi_1$, which represents the fraction of super-Earths at short orbital periods, increases with stellar mass. This behavior can be explained if the relative amount of sub-Neptunes evaporated to super-Earth sizes increases with stellar mass as discussed in Section (\ref{sec:atm}). Within the $1\sigma$ uncertainties, the fraction $\chi_2$ appears to be independent of stellar mass, suggesting that at long periods, the relative fraction of super-Earths is constant with stellar mass.

\section{Discussion} \label{sec:discussion}

In this work, we analyse the population of small ($1- 3.5\,\EarthRad$), close-in ($2-100$\,days) planets and their dependence on stellar mass. We show a trend in which super-Earth-sized planets are more common than sub-Neptunes out to a certain orbital period, beyond which the opposite is true. When this trend is fit independently across several bins of host star mass, we find evidence that the orbital period where super-Earths and sub-Neptunes are equally abundant moves outward with increasing stellar mass. Furthermore we find that, within this transition, the fraction of small planets comprised of super-Earths increases around heavier stars. However, outside this transition, the relative fraction of super-Earths appears to be independent of stellar mass.

\subsection{Connection to Atmospheric Loss Mechanisms} \label{sec:atm}
A common explanation for the existence of the radius valley and the bimodal size distribution of small planets is atmospheric mass loss. Because the dependence between planet radius and envelope mass is strongly nonlinear (e.g., \citealp{Lopez&Fortney2014}, Fig.~13 in \citealp{Fulton2017}), a sub-Neptune-size planet ($\sim 2.5$\,R$_\oplus$) with just $0.5\%$ of its total mass in a H/He envelope could shrink in size down to a super-Earth ($\sim 1.5$\,R$_\oplus$) if it were to lose this atmospheric envelope.

There are two prevalent mechanisms of atmospheric loss which differ in the source of energy driving hydrodynamic escape. Photoevaporation relies on high energy X-ray and EUV (also known as XUV) photons from the host star to heat the upper atmosphere \citep{Lammer2003, Lopez&Fortney2013, Owen&Wu2013, Jin2014}. In contrast, core-powered mass loss involves an outflow of the upper atmosphere driven by a mix of the planet core's residual heat from the accretion phase and the bolometric luminosity of the star \citep{Ginzburg2016, Ginzburg2018, GuptaSchlichting2019, GuptaSchlichting2020}. In the former, a planet's mass loss rate is assumed to be directly proportional to the XUV luminosity of its host star; in the latter, the mass loss rate is dependent on the planet's equilibrium temperature which (partially) depends on the bolometric luminosity of the star.

Both XUV and bolometric luminosity increase with stellar mass, so because more massive stars provide higher relevant flux than low mass stars for a given orbital separation, both models predict that more atmospheric mass can be lost to farther orbital periods around heavier, more luminous stars. In turn, the limit where super-Earths no longer appear more abundant than sub-Neptunes should increase with stellar mass in both photoevaporation and core-powered mass loss, in agreement with what is observed in Figure~(\ref{fig:Mass-vs-Turnover}).

\begin{figure*}
    \centering
    \includegraphics[width=0.8\textwidth]{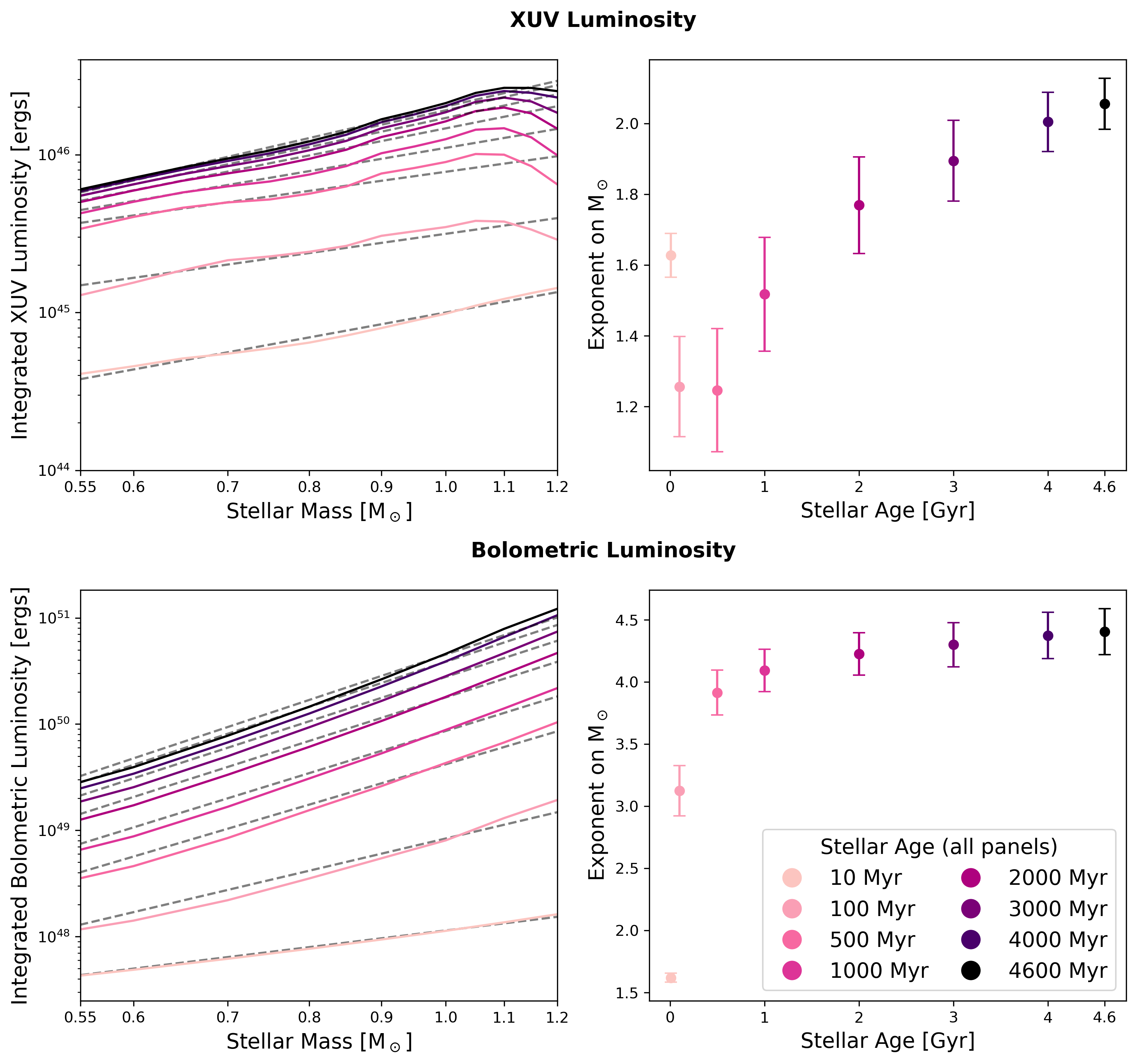}
    \caption{Evolutionary tracks and stellar mass dependence for XUV and bolometric luminosity. Colors and corresponding ages are the same for all panels. \textbf{Top Left:} Time-integrated cumulative XUV luminosity (relevant for photoevaporation) for a selection of key ages, including the approximate age \citep[$\sim4.6$\,Gyr, ][]{Berger2020a} of the \kepler field, using evolutionary tracks from \citet{Johnstone2021}. Grey dashed lines indicate log-log relations fit independently at each time. \textbf{Top Right:} The exponential dependence $\xi$ of integrated XUV luminosity on stellar mass for the log-log fits of $L \propto M_*^\xi$ as a function of stellar age. \textbf{Bottom Left:} Similar to the XUV luminosity panel, but now using bolometric evolutionary tracks (relevant for core-powered mass loss) using data from \citet{BHAC2015}. \textbf{Bottom Right:} Similar to the integrated XUV exponent panel, but now using integrated bolometric luminosity.}
    \label{fig:Luminosity}
\end{figure*}

The dependence on stellar mass presented in this work is difficult to contextualize, because neither model has an easy proxy for the population-level metric of a transition period as used here. However, it is still useful to consider how the relevant luminosities scale with stellar mass. Because we are interested in how the cumulative effects of atmospheric loss have led to the modern population of exoplanets and trends therein, we focus on the cumulative or time-integrated luminosity. For photoevaporation which relies on XUV luminosity, we employ the evolutionary tracks of \citet{Johnstone2021} which are based on empirical relations between stellar parameters and high energy emission. We then do a log-log fit to measure how the cumulative XUV luminosity changes with stellar mass, and how this dependence evolves with system age. The tracks themselves are shown in the upper panels of Figure~(\ref{fig:Luminosity}), along with approximate values for the exponential dependence of the XUV luminosity on stellar mass (i.e., the value of exponent $\xi$ in a relationship for $L_{XUV} \propto M_*^\xi$). We repeat a similar process for the dependence on bolometric luminosity, which is relevant for core-powered mass loss, using the tracks of \citet{BHAC2015}, displayed in the lower panels of Figure~(\ref{fig:Luminosity}). At the modern age of the \kepler field - reported as $4.58$\,Gyr in \citet{Berger2020a}, which we approximate as $4.6$\,Gyr due to resolution limits in the adopted luminosity tracks - we note that $L_{XUV} \propto M_*^{2.0 \pm 0.1}$ and $L_{BOLO} \propto M_*^{4.4 \pm 0.2}$ based on this simple characterization.

It is important to note that \citet{Johnstone2021} provide XUV luminosity tracks for different stellar initial rotation rates based on percentiles of an observed rotation distribution for stars with an age of $150$\,Myr. Because the differences in these rotation rates are most pronounced for stellar masses lower than what is used in this work (i.e., $M_* < 0.56\,\SolarMass$), we find no significant change when comparing the 16th, 50th, and 84th rotation percentiles with time. Thus we employ only the 50th percentile (or ``medium" rotation rate) stars, but note that considerations of stellar rotation rates may be important for detailed modeling efforts in the future, especially for planets orbiting lower mass stars.

In light of these mass-luminosity relations, we consider a scenario where the transition between super-Earths and sub-Neptunes occurs at a fixed incident flux. For a planet orbiting a star of luminosity $L_*$ at semi-major axis $a$, the incident flux $I$ on the planet can be written as:
\begin{equation}
    I \propto L_* a^{-2}.
\end{equation}
Kepler's third law allows us to relate this flux to an orbital period $P$ for a planet around a star of mass $M_*$:
\begin{equation}
    a^3 \propto M_* P^2 \longrightarrow I \propto L_* M_*^{-2/3} P^{-4/3}.
\end{equation}
If the relevant luminosity (wavelength-ambiguous for generality) scales with stellar mass following $L_* \propto M_*^\xi$, then the flux would scale as:
\begin{equation}
    I \propto M_*^{\xi - 2/3} P^{-4/3}.
\end{equation}
Adopting two planets at the same flux $I_1 = I_2$ but orbiting different stars $M_{*,1}$ and $M_{*,2}$ allows us to write:
\begin{equation}
    \frac{I_1}{I_2} = \left(\frac{M_{*,1}}{M_{*,2}}\right)^{\xi - 2/3} \left(\frac{P_1}{P_2}\right)^{-4/3} = 1
\end{equation}
\begin{equation}
    \frac{P_1}{P_2} = \left(\frac{M_{*,1}}{M_{*,2}}\right)^{\frac{3}{4} \cdot \left[\xi - \frac{2}{3}\right]},
\end{equation}
where we note that these orbital periods correspond to the fixed flux where this transition occurs. Thus, we would expect the orbital period of this transition to follow a relationship with stellar mass like $P_\mathrm{trans} \propto M_*^\alpha$ where $\alpha = \frac{3}{4}\cdot\left[\xi-\frac{2}{3}\right]$.

In this work we observe $\alpha = 1.7\pm0.2$, corresponding to a mass-luminosity relation with $\xi = 2.9\pm0.3$. This is distinct from both the XUV ($\xi=2.0 \pm 0.1$) and bolometric ($\xi=4.4 \pm 0.2$) luminosity relations presented in Figure~(\ref{fig:Luminosity}) for the median age of the \kepler field (predicting $\alpha=1.0\pm0.1$ or $\alpha=2.8\pm0.1$, respectively). One caveat of this simple comparison is that atmospheric loss is time-dependent, as both host star luminosity and planetary conditions will evolve with system age. The novel discrepancy between observations and predictions from either model may be resolved with more detailed time-dependent modeling efforts beyond the scope of this work. We advocate that this orbital period of transition with a dependence on stellar mass (and therefore luminosity) may serve as a useful metric by which to compare modeling efforts with the \kepler sample. Additionally, $P_\mathrm{trans}$ could provide a useful constraint by which to distinguish (or reconcile) the photoevaporation and core-powered mass loss models in the future.


\subsection{Implications for Small Planets in the Habitable Zone} \label{sec:HZ}

The frequency of Earth-sized habitable zone planets is a valuable science result for population studies, as it is a critical input to determine the yield of future missions aiming to discover and characterize Earth analogues. A typical metric for habitable zone exoplanet studies is $\eta_\oplus$, or the occurrence rate of Earth-sized ($\approx 1 \, \EarthRad$) planets within the habitable zone. Because the planet radius range defining ``Earth-sized" and the orbital period range of the habitable zone differ across the community, the ExoPAG Study Analysis Group 13\footnote{\url{https://exoplanets.nasa.gov/exep/exopag/sag/\#sag13}} suggests the use of the differential (normalized) occurrence rate $\Gamma_\oplus$, defined as:
\begin{equation}\label{eqn:Gamma}
    \Gamma_\oplus = \frac{\partial^2 N}{\partial \ln \Porb \partial \ln \Rp} \bigg\rvert_{\Porb = 1\,\mathrm{yr}, \Rp = \EarthRad}.
\end{equation}
This is akin to calculating $\eta_\oplus$ in some orbital period and radius bin, then dividing the occurrence by the dimensions of the bin in units of natural log Earth years and radii. 

As mentioned in Section~(\ref{sec:intro}), the lack of long-period reliable detections with \kepler means that typical completeness-weighted occurrence rate calculations are not applicable to the habitable zone. Instead, many works opt to describe the behavior of exoplanet populations in the short-period regime (where there are numerous detections) and extrapolate these functions out to longer habitable zone orbital periods. However, failing to consider the over-representation of Earth-sized planets at short orbital periods due to atmospheric loss may lead to overestimates in the occurrence of Earth-sized, longer-period planets \citep{Lopez&Rice2018, Pascucci2019}.

In this work, we expand upon previous studies of small planet habitable zone occurrence by specifically accounting for the effects of remnant cores at short orbital periods. Because our fractional occurrence model probes the interchange between super-Earth and sub-Neptune populations with orbital period, it also serves to separate two physically distinct regimes: the short-period $\left(\Porb < P_\mathrm{trans}\right)$ regime where atmospheric loss converts a large fraction of sub-Neptunes into super-Earths, and the long-period $\left(\Porb > P_\mathrm{trans}\right)$ regime where atmospheric loss is inefficient. The latter is presumably representative of all orbital periods beyond both $P_\mathrm{trans}$ and the $100$\,day cut-off modeled in this work, barring an increase in the relative number of super-Earths at longer orbital periods which cannot be discerned from current data.

Therefore, we argue that the asymptotic fraction $\chi_2$ (which the fractional occurrence plateaus towards even within $100$\,days; see Figure~\ref{fig:FracOcc}) is a suitable constraint for measuring how much of the small planet population is comprised of super-Earths at longer orbital periods. Because this fraction is apparent in the regime where atmospheric loss is ineffective, it may represent an underlying population of intrinsically rocky super-Earths present at all periods. In this sense, the short-period fraction $\chi_1$ could consist of at least two groups: a fraction $\left(\chi_1 - \chi_2\right)$ of planets which have lost their atmospheres, and a fraction $\chi_2$ contributed by the true rocky super-Earths ($\sim33\%$, independent of stellar mass; see Table~\ref{tab:fits}). This interpretation is supported by the work of \citet{Neil2020}, in which an evaluation of various mixture models favors a combination of planets shaped by envelope mass loss alongside an independent population of rocky planets. We note that while $\chi_1$ increases with stellar mass and $\chi_2$ does not (see bottom row of Figure~\ref{fig:BestFit-vs-Mstar}), we do not find any clear stellar mass dependence in the fraction of planets which have lost their atmospheres (e.g., the distribution of $\chi_1 - \chi_2$ peaks in the $\left[0.91,1.01\right]\,\SolarMass$ bin with $\sim56\%$) due to larger variations in the estimated values of $\chi_2$.

The long-period fraction is insightful, but is not particularly useful for occurrence calculations without also knowing what the overlaying (non-fractional) occurrence of small planets is at a given orbital period. Our model also includes this distribution (via Equation~\ref{eqn:bpl}), such that we may first assess the occurrence of small planets via a broken power law, then further divide this between super-Earths and sub-Neptunes with our fractional occurrence model, as a function of orbital period. Because the behavioural shifts at $P_\mathrm{break}$ and $P_\mathrm{trans}$ occur while still within the close-in population, we conclude that the long-period components of our model are suitable for extrapolation to longer orbital periods - most notably, the habitable zone.

\begin{deluxetable*}{cccccccc}
\tabletypesize{\footnotesize}
\tablecolumns{6}
\tablewidth{0pt}
    \tablecaption{Habitable Zone Occurrence\label{tab:gamma}}
    \tablehead{
    \colhead{Stellar Mass Range} & \colhead{Habitable Zone} & \colhead{$\Gamma_\mathrm{sE}$} & \colhead{$\Gamma_\mathrm{sN}$} & \colhead{$\eta_\mathrm{sE+sN}$} & \colhead{$\Gamma_\mathrm{sE+sN}$} & \colhead{$\eta_\oplus$} & \colhead{$\Gamma_\oplus$} \vspace{-0.15cm} \\
    \colhead{$\left[M_\odot\right]$} & \colhead{[days]} &
    \multicolumn{6}{c}{[\%]}
    }
    \startdata
    0.56 - 0.81 & 151 - 360 & ${13.29}_{-4.80}^{+6.95}$ & ${23.29}_{-8.55}^{+11.06}$ & ${20.43}_{-6.60}^{+8.94}$ & ${18.75}_{-6.06}^{+8.20}$ & ${8.80}_{-3.18}^{+4.61}$ & ${13.28}_{-4.80}^{+6.95}$ \\
    0.81 - 0.91 & 276 - 629 & ${28.52}_{-9.45}^{+11.51}$ & ${56.50}_{-18.73}^{+20.60}$ & ${44.39}_{-14.52}^{+13.64}$ & ${42.93}_{-14.04}^{+13.19}$ & ${17.93}_{-5.95}^{+7.24}$ & ${28.51}_{-9.46}^{+11.51}$ \\
    0.91 - 1.01 & 370 - 826 & ${11.28}_{-4.27}^{+6.23}$ & ${40.63}_{-12.46}^{+15.15}$ & ${25.22}_{-7.54}^{+9.03}$ & ${25.11}_{-7.51}^{+8.99}$ & ${6.88}_{-2.61}^{+3.81}$ & ${11.26}_{-4.27}^{+6.24}$ \\
    1.01 - 1.16 & 497 - 1088 & ${5.95}_{-2.52}^{+4.68}$ & ${18.45}_{-7.74}^{+10.55}$ & ${11.33}_{-4.44}^{+6.74}$ & ${11.56}_{-4.53}^{+6.87}$ & ${3.54}_{-1.50}^{+2.79}$ & ${5.94}_{-2.52}^{+4.67}$ \\
    1.16 - 1.63 & 859 - 1822 & ${24.95}_{-8.75}^{+14.72}$ & ${64.36}_{-21.71}^{+35.54}$ & ${38.59}_{-13.02}^{+19.39}$ & ${40.94}_{-13.81}^{+20.57}$ & ${14.28}_{-5.00}^{+8.44}$ & ${24.91}_{-8.72}^{+14.72}$ \\
    \hline
    0.56 - 1.63 & 363 - 811 & ${15.30}_{-4.05}^{+5.56}$ & ${33.27}_{-8.70}^{+11.90}$ & ${23.16}_{-5.83}^{+8.23}$ & ${22.98}_{-5.79}^{+8.16}$ & ${9.37}_{-2.48}^{+3.40}$ & ${15.29}_{-4.05}^{+5.55}$ \\
    \enddata
    \tablecomments{Habitable zone occurrence rates are computed by extrapolating our distribution function in orbital period to the maximum and moist greenhouse boundaries as defined in \citet{Kopparapu2013}. Because $\Gamma$ is normalized by the width of the relevant natural log radius bin, $\Gamma_\mathrm{sE+sN}$ (corresponding to the radius range of $R_p = 1 - 3.5\,\EarthRad$) is smaller than the sum of $\Gamma_\mathrm{sE}$ and $\Gamma_\mathrm{sN}$, which are both calculated for smaller radius ranges. $\eta_\oplus$ and $\Gamma_\oplus$ are calculated for the conventional Earth-sized range of $R_p = 0.7 - 1.5\,\EarthRad$. This involves extrapolation for radii $R_p < 1.0\,\EarthRad$ which are not included in the fitting range of this work.}
\end{deluxetable*}

For each bin of stellar mass, we compute the average stellar effective temperature \citep[using values from][]{Berger2020a} and use this to estimate the orbital period range of the habitable zone corresponding to the maximum and moist greenhouse boundaries as defined in \citet{Kopparapu2013}; the orbital period bounds are included in Table~(\ref{tab:gamma}). We then extrapolate Equation~(\ref{eqn:df}), integrating in orbital period from the inner- to outer-edge of the habitable zone, and separately integrate over three radius regimes: super-Earths ($1\,\EarthRad \leq \Rp \leq R_\mathrm{valley}$), sub-Neptunes ($R_\mathrm{valley} \leq \Rp \leq 3.5\,\EarthRad$), and the combination of the two ($1 \leq \Rp \leq 3.5\,\EarthRad$). The results are then normalized by the period-width of the habitable zone (in years) and radius-width of the relevant size regime to provide estimates for $\Gamma_\mathrm{sE}$, $\Gamma_\mathrm{sN}$, and $\Gamma_\mathrm{sE+sN}$, respectively, in Table~(\ref{tab:gamma}). We also provide $\eta_\mathrm{sE+sN}$ (i.e., the occurrence before normalizing by integration ranges), from which $\eta_\mathrm{sE}$ and $\eta_\mathrm{sN}$ can be approximately recovered by multiplying with $\chi_2$ and $\left(1-\chi_2\right)$, respectively. This is not the case for $\Gamma$, as the radius range used to normalize the occurrence varies between classifications.

Additionally, we follow the method outlined above to estimate the habitable zone occurrence of Earth-sized planets (conventionally listed as $0.7 \leq \Rp \leq 1.5\,\EarthRad$). We note that the range of radii used in this work extends only down to $R_p = 1.0\,\EarthRad$, so to evaluate the Earth-sized regime, we must extrapolate our occurrence function down to smaller radii. To do this, we assume a marginalized radius distribution that is constant in $\log{\Rp}$ for $R_p < R_\mathrm{valley}$, consistent with the occurrence rate calculations of both \citet{FultonPetigura2018} and our work with \epos{} (see Figure~\ref{fig:MarginalizedRadiusDist}; Appendix~\ref{sec:pldf}). After extrapolating both out in orbital period and down in radius, we integrate across the habitable zone to provide both the exact $\eta_\oplus$ and normalized $\Gamma_\oplus$ for each of our five stellar mass bins in Table~(\ref{tab:gamma}). The distinction between $\Gamma_\oplus$ and $\Gamma_\mathrm{sE}$ persists mostly as a formalism to clarify that the former properly includes smaller planets down to $0.7\,\EarthRad$, and because $\Gamma_\oplus$ is more conventional for occurrence rate studies (as discussed in Section~\ref{sec:GammaLit}). We note that both groups are subjected to the same general distribution - therefore providing comparable trends in $\eta$ - and that normalizing by different $\dd\ln{\Rp}$ widths coincidentally produces near-identical values of $\Gamma$.

The variation in $\Gamma_\oplus$, $\Gamma_\mathrm{sE}$, $\Gamma_\mathrm{sN}$, and $\Gamma_\mathrm{sE + sN}$ with stellar mass follows closely that of the power law slope $\beta_2$ in orbital period (see Figure~\ref{fig:BestFit-vs-Mstar}). This implies that the uncertainty and variation in $\Gamma_\oplus$ is dominated by the uncertainty in how the overall orbital period distribution extends to long periods, more so than any other component of our model (such as the long-period super-Earth fraction). The poor constraints on $\beta_2$ in each bin are largely driven by a small number of long-period detections, as combining the entire FGK sample into a single fit - thereby increasing the number of detections - provides tighter constraints on both $\beta_2$ and $\Gamma_\oplus$; see Table~(\ref{tab:fits}). To assess if the increased precision on $\Gamma_\oplus$ from broader bins may come at the cost of smoothing over any trends with stellar mass (e.g., the average number of planets per star $F_0$), we repeat the procedure outlined in Section~(\ref{sec:epos}) using three stellar mass bins (instead of one or five). We find $\Gamma_\oplus = \{18.51^{+6.65}_{-4.63}, 11.46^{+4.44}_{-3.54}, 10.39^{+7.16}_{-4.44} \}\%$ for bins with boundaries $\{0.56, 0.88, 1.05, 1.63\}\,\SolarMass$, all consistent with that of the combined FGK sample. We affirm that the uncertainty on $\beta_2$ and thus $\Gamma_\oplus$ is smallest when the sample of long-period planets is largest, and note that the \kepler detected population of small, long-period planets is presently too sparse to confidently assess any trends with stellar mass in the habitable zone.

\begin{figure*}
    \centering
    \includegraphics[width=0.8\textwidth]{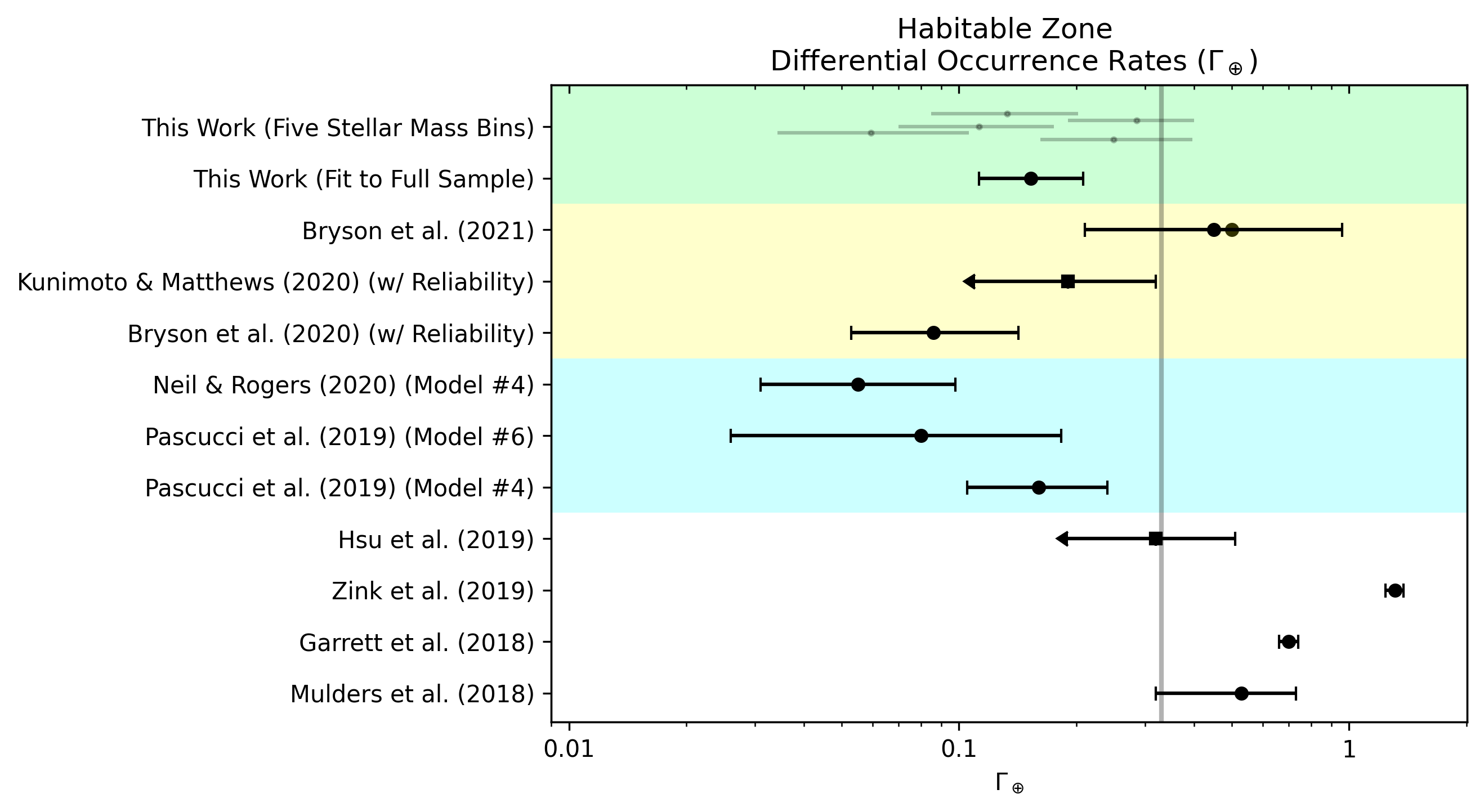}
    \caption{Comparison of habitable zone occurrence rates for Earth-sized planets. Our result for the combined $[0.56, 1.63]\,\SolarMass$ sample is plotted at full opacity, and lower-opacity points denote results from individual fits to five stellar mass bins with boundaries spanning $\{0.56, 0.81, 0.91, 1.01, 1.16, 1.63\}\,\SolarMass$ (increasing in mass from top to bottom). Circles indicate values assuming the occurrence follows a functional form (our work, \citealp{Bryson2021, Bryson2020, Neil2020, Pascucci2019, Zink2019, Garrett2018}, and \citealp{Mulders2018}), and squares indicate values from grid-based occurrence rates \citep{Kunimoto&Matthews2020, Hsu2019}. Left-pointing arrows indicate upper limits. The blue region represents works directly considering an underlying population of intrinsically rocky planets, the yellow region represents works considering reliability, and our work (shown in green) implements both considerations. The vertical grey line indicates the occurrence values from ExoPAG SAG13 via \citet{Kopparapu2018} used in the final reports of the LUVOIR and HabEx missions. A similar plot for $\Gamma_\oplus$ estimates prior to \citet{Bryson2021} appears in \citet{Kunimoto&Matthews2020} (their Figure 14).}
    \label{fig:HZ_Comparison}
\end{figure*}

\subsubsection{Comparisons to Previous Works}\label{sec:GammaLit}

Our result for $\Gamma_\oplus$ from the combined FGK sample and five stellar mass bins are compared with other $\Gamma_\oplus$ values from recent literature in Figure~(\ref{fig:HZ_Comparison}). We exclude works prior to 2018 which did not include the full \kepler \texttt{DR25} catalog \citep{Thompson2018}. The combined FGK value of $\Gamma_\oplus \approx 15_{-4}^{+6}\%$ reported in this work is consistent at the $1\sigma$ level with \citet{Bryson2021}, \citet{Bryson2020}, and \citet{Pascucci2019}, and is not excluded by the upper limits of \citet{Kunimoto&Matthews2020} or \citet{Hsu2019}. As suggested by \citet{Lopez&Rice2018}, works adopting independent broken power laws in orbital period and radius (e.g., \citealp{Zink2019, Mulders2018}) may be overestimating $\eta_\oplus$ due to atmospheric loss effects. Predictions by \citet{Zink2019} seem particularly large by comparison, which is attributable to their considerations of how transit multiplicity can lead to a detection-order-based decrease in detection efficiency, and perhaps additionally a lack of reliability treatment.

The estimate of \citet{Kunimoto&Matthews2020} comes from an occurrence evaluated in a $\left[237,500 \right]$\,day orbital period bin and extrapolated to $860$\,days (the outer edge of their habitable zone). The agreement with values reported in this work suggests that the small planet population within $100$\,days (where \kepler detections are numerous) may indeed be a suitable indicator for the longer-period planet population (where detections are sparse) \textit{when implementing proper considerations of both reliability and trends in fractional occurrence}. However, simplifying the latter with a flat orbital period cut for all FGK stars may not be suitable due to the strong stellar-mass dependence of $P_\mathrm{trans}$ described in Section~(\ref{sec:ParameterDetails}). For example, \citet{Pascucci2019} show that $\Gamma_\oplus$ estimates may be sensitive to the minimum fitted orbital period by considering cuts at $12$ and $25$\,days (their models $4$ and $6$, respectively). Thus, a more robust consideration of fractional occurrence trends (e.g., the model used in this work) may be necessary when including any of the close-in population for extrapolations to longer orbital periods.

Noting the sensitivity of habitable zone extrapolations to the value of orbital period power law exponents, \citet{Bryson2021} characterize occurrence in terms of instellation flux rather than orbital period. The flux-based approach allows for more direct inclusions of observed planets in the habitable zones of more stars (while an orbital period approach might require unbounded extrapolations for some stars with no reliable detections in long-period regimes). As derived in \citet{Bryson2021}, a power law in orbital period with slope $\beta$ has a corresponding flux slope of approximately $\nu = -\frac{3}{4}\left(\beta + \frac{7}{3}\right)$, which is accompanied by a stellar temperature-dependent term when converting from $\dd f / \dd P$ to $\dd f / \dd I$. The $1\sigma$ range of $\beta_2$ values presented in Table~(\ref{tab:fits}) spans a range of $-0.78 < \nu < -1.14$, consistent with the findings of \citet{Bryson2021} for the flux-dependent power law component of their model(s). As a comparison, \citet{Bryson2020} adopt a single power law in orbital period (fit down to $50$\,days, beyond the break in our broken power law) with a typical exponent of $\beta \approx -0.8\pm0.2$ ($\nu\approx-1.15\pm0.15$), consistent with all values of $\beta_2$ found in this work.

Given the consistency in long-period / low-instellation occurrence rate slopes between works, an additional factor is needed to explain the wide range in habitable zone differential occurrence rate estimates when calculated in orbital period versus instellation flux - though (as Bryson et al. note) it remains to be shown if $\Gamma_\oplus$ is equivalent in these regimes. One explanation may be that the instellation flux approach is biased towards shorter orbital periods where there are more \kepler planet detections, corresponding to lower-mass stars where the habitable zone is closer-in. If lower mass (K-type) stars have a higher number of small planets per star (e.g., $F_0$, see Figure~\ref{fig:BestFit-vs-Mstar}), such a bias could cause inflated predictions of $\Gamma_\oplus$ for bounded extrapolations to hotter stars.

On the subject of instellation, recent work by \citet{Loyd2020} suggests that the stellar mass-dependent scaling of the radius valley by \citet{Wu2019} adopted in this work may vanish when correcting for differences in instellation flux. Adopting an instellation-adjusted radius valley persistent at $1.8\,\EarthRad$ for all relevant stellar masses would cause fewer planets to be labelled as super-Earths around heavier stars (for which we have adopted $R_\mathrm{valley} > 1.8\,\EarthRad$; see Table~\ref{tab:MassBins}). This would lead to a decrease in the super-Earth fractions $\chi_1$ and $\chi_2$ --- and by extension of the latter, a decrease in $\Gamma_\oplus$ --- which would be more pronounced for heavier stars. In any case, we affirm that estimates of $\eta_\oplus$ and $\Gamma_\oplus$ are sensitive to the adopted parameterization.

Our interpretation of the population described by $\chi_2$ as a group of intrinsically rocky planets is supported by the calculations of \citet{Neil2020}. When considering a mixture model\footnote{Values cited are from Model $\#4$ of Table $2$ in \citet{Neil2020}, which the authors suggest better matches the radius distribution for $R_p < 1.5\,\EarthRad$.} with components for both an intrinsically rocky population and planets which experience atmospheric loss, they find an occurrence rate of $\Gamma_\oplus \approx 5.6_{-2.5}^{+4.4}\%$. Their sample of stars uses a more truncated stellar mass range ($4700 < T_\mathrm{eff} < 6500$\,K, or approximately $0.75 < M_* < 1.33\,\SolarMass$; \citealp{MamajekTable}), for which the average stellar mass falls near the boundary of our $\left[0.91, 1.01\right]$ and $\left[1.01, 1.16\right]\,\SolarMass$ bins. As these bins are most likely to be reminiscent of the population studied by \citet{Neil2020}, it is encouraging that we find consistent values of $\Gamma_\oplus = 11.3^{+6.2}_{-4.3}\%$ and $5.9^{+4.7}_{-2.5}\%$. Their $\Gamma_\oplus$ prediction does not change significantly when further limiting their occurrence calculation to within $50\%$ of Earth's mass, suggesting that at habitable zone orbital periods (a) our extrapolation down in planet radius and (b) the interpretation of $\chi_2$ as representing a true rocky population are consistent with higher-dimensional approaches incorporating planet mass.


\subsubsection{Impact on Future Missions}\label{sec:decadal}

It is useful to discuss our value of $\Gamma_\oplus$ for the combined FGK sample in the context of the consensus study ``Pathways to Discovery in Astronomy and Astrophysics for the 2020s" (\citealp{Decadal}, hereafter ``decadal survey"). In light of two mission concept studies - the Large Ultraviolet / Optical / Infrared Surveyor (LUVOIR; \citealp{LUVOIR}) and the Habitable Exoplanet Observatory (HabEx; \citealp{HabEx}) - the decadal recommends a $\sim 6$\,m-class telescope for exoplanet detection and characterization. This recommendation is not dissimilar to the proposed LUVOIR-B, which had an inscribed diameter of 6.7m (and a circumscribed diameter of 8m), and thus we expect that the two would produce comparable yields.

The proposed sample of $\sim25$ potentially habitable exoplanets stems from an occurrence rate of $\eta_\oplus = 24\%$ for planets of radius $[0.6, 1.4]$\,$\EarthRad$ with semi-major axes of $[0.95, 1.67]$\,AU around Sun-like stars (\citealp{LUVOIR, HabEx}; taken from the ExoPAG SAG13 synthesis of \citealp{Kopparapu2018}.) This equates to a differential occurrence rate of $\Gamma_\oplus = 33\%$ (grey line in Figure~\ref{fig:HZ_Comparison}) which is larger than most predictions in recent literature, including this work. A re-scaling of LUVOIR-B's predicted mission yield for Earth-sized habitable zone planets ($28_{-17}^{+30}$) by our differential occurrence rate for the combined FGK sample suggests a smaller yield of $13_{-1}^{+2}$ such planets. Following the values presented in Table~(\ref{tab:gamma}), we would predict a comparable yield of super-Earths and roughly twice as many sub-Neptunes (the dominant species of long-period small planets). It will be interesting to explore the habitability of these larger planets at relevant separations. Finally, surveys targeting single-star systems could see a factor of $\sim2$ increase in $\Gamma_\oplus$ by avoiding the formation-suppressing and transit depth-diluting effects of binary systems \citep{Moe&Kratter2021}.


\section{Summary and Conclusions} \label{sec:summary}

In this work, we explore the occurrence of super-Earths $\left(\sim 1-2\,\EarthRad\right)$ and sub-Neptunes $\left(\sim2-3.5\,\EarthRad\right)$ as a function of orbital period in several stellar mass bins spanning $0.56 - 1.63\,\SolarMass$ and corresponding to FGK-type stars. 

\begin{itemize}
    \item We confirm evidence of a transition in the occurrence rates between two regimes of close-in \kepler planets: where super-Earths are more abundant at shorter orbital periods, and where sub-Neptunes are more abundant at longer orbital periods.

    \item Using a parametric model including a metric of fractional occurrence, we find that the orbital period beyond which super-Earths are no longer more common than sub-Neptunes moves further out around heavier, more luminous stars. This period follows a relation like $P_\mathrm{trans} \propto \Mstar^{1.7 \pm 0.21}$, increasing from $P_\mathrm{trans} \approx 8_{-1}^{+1}$\,days around $M_* = 0.56-0.82\,\SolarMass$ stars to $P_\mathrm{trans} \approx 27_{-7}^{+18}$\,days around $M_* = 1.16-1.63\,\SolarMass$ stars.
    
    \item We find a stellar mass dependence for the fraction of short-period planets which are super-Earth-sized. This fraction $\chi_1$ increases from $0.75_{-0.05}^{+0.05}$ to $0.87_{-0.05}^{+0.05}$ while increasing stellar mass from $M_* = 0.56-0.82\,\SolarMass$ to $M_* = 1.16-1.63\,\SolarMass$.
    
    \item We extrapolate our parametric model to report the differential occurrence of Earth-sized planets in the habitable zone of FGK stars as $\Gamma_\oplus = 15^{+6}_{-4}\%$. \kepler's poor sampling of small planets at long orbital periods is the primary source of uncertainty preventing the resolution of any discernible trends with $\Gamma_\oplus$ on stellar mass.
\end{itemize}

The evidence of an orbital period-dependent transition positively correlated with stellar mass (and therefore luminosity) is predicted by both models of photoevaporation and core-powered mass loss. This supports the idea that the population of close-in planets is sculpted by atmospheric loss. However, more detailed work is needed to discern which mechanism(s) can properly recover the observed trend in $P_\mathrm{trans}$ presented here. The increased fraction $\chi_1$ with stellar mass suggests that higher planetary atmospheric mass loss around F-type stars produces more close-in super-Earths. This does not lead to more Earth-sized habitable zone planets because the strength of the mechanism(s) driving atmospheric loss falls off with orbital period. Thus, models of atmospheric loss motivate a treatment accounting for the presence of remnant cores at close-in orbital periods, when using the close-in population to extrapolate out towards the habitable zone. Such extrapolations are in good agreement with works fitting the orbital period distribution of longer-period planets supplemented with considerations of planet reliability. 

The range of occurrence rates from recent works illustrates that our knowledge of $\eta_\oplus$ remains limited based on current data, and we note that our interpretations are further limited by our understanding of planet formation and evolution. The true value of $\eta_\oplus$ could be larger than what is presented here if the population of $0.7-1\,\EarthRad$ planets increases with orbital period. However, this cannot be assessed using current data due to low completeness and reliability for planets smaller than Earth at long orbital periods. Thus, more reliable estimates of $\eta_\oplus$ will require a better determination of the overall distribution of small planets at long orbital periods, and their further distribution with stellar mass.

Future works exploring the features and trends presented here, in both the \kepler dataset and beyond (e.g., \textit{K2}), will be invaluable to resolving the many open questions in exoplanets today. The continuing efforts of the TESS mission and team will serve to better refine our understanding of possible stellar mass dependencies by expanding the baseline of host star masses. Recent TESS discoveries have already begun to populate the hot-Neptune desert and radius valley with young planets, so the targeting of younger ($\sim$\,Myr) systems will allow us to probe an era before atmospheric loss fully molds the exoplanet population. The recent launch of JWST \citep{JWST} brings a new tool with which to characterize planetary atmospheres for a range of ages and planet radii, further clarifying how the mechanism(s) of atmospheric mass loss cause close-in planets to evolve with time.

At present, a major focus of astrobiology and the exoplanet community is to find Earth-sized planets in the habitable zones of Sun-like stars. An additional goal is to search for and characterize biosignatures in the atmospheres of these Earth analogues. The high-resolution requirement of studying biosignatures means that this process can only be carried out for nearby planets, yet - following the results of this work - a $\sim6$\,m-class telescope surveying 100 nearby stars would likely lend only $\sim 13$ Earth-sized habitable zone planets for characterization. Furthermore, there are presently few comprehensive lists or metrics for which nearby stars might host these target planets. To produce realistic mission yields and informed target selection criteria, it is thus imperative to understand both the occurrence rate $\eta_\oplus$ \textit{and} how this occurrence may change with host star properties, including mass. Future missions should continue to account for uncertainties in $\eta_\oplus$ estimates to verify their science cases are resilient to pessimistic outcomes, or otherwise advocate for future statistical missions to reduce these uncertainties. We are excited by the exoplanet community's fervent efforts to understand, constrain and implement habitable zone occurrence rates while working towards the next generation of exoplanet science. 

\software{\texttt{NumPy} \citep{numpy}, \texttt{SciPy} \citep{scipy}, \texttt{Matplotlib} \citep{pyplot}, \texttt{emcee} \citep{ForemanMackey2012}, \texttt{corner} \citep{corner}, \texttt{epos} \citep{Mulders2018}}


\section{Acknowledgements}

We would like to thank the anonymous referee for their insightful feedback regarding the adopted parameterization of our fractional occurrence model.
I.P., G.B, and R.B.F. acknowledge support from the NASA Astrophysics Data Analysis Program under Grant No. 80NSSC20K0446.
G.D.M. acknowledges support from ANID --- Millennium Science Initiative --- ICN12\_009.
This material is based upon work partly supported by NASA under Agreement No. NNX15AD94G for the program ``Earths in Other Solar Systems" and under Agreement No. 80NSSC21K0593 for the program ``Alien Earths”. The results reported herein benefited from collaborations and/or information exchange within NASA’s Nexus for Exoplanet System Science (NExSS) research coordination network sponsored by NASA’s Science Mission Directorate.
\bibliographystyle{apj}
\bibliography{ref}

\begin{thebibliography}{}
\expandafter\ifx\csname natexlab\endcsname\relax\def\natexlab#1{#1}\fi

\bibitem[{{Baraffe} {et~al.}(2015){Baraffe}, {Homeier}, {Allard}, \&
  {Chabrier}}]{BHAC2015}
{Baraffe}, I., {Homeier}, D., {Allard}, F., \& {Chabrier}, G. 2015, \aap, 577,
  A42

\bibitem[{{Bean} {et~al.}(2021){Bean}, {Raymond}, \& {Owen}}]{Bean2021}
{Bean}, J.~L., {Raymond}, S.~N., \& {Owen}, J.~E. 2021, Journal of Geophysical
  Research (Planets), 126, e06639

\bibitem[{{Beaug{\'e}} \& {Nesvorn{\'y}}(2013)}]{Beauge2013}
{Beaug{\'e}}, C., \& {Nesvorn{\'y}}, D. 2013, \apj, 763, 12

\bibitem[{{Berger} {et~al.}(2020{\natexlab{a}}){Berger}, {Huber}, {Gaidos},
  {van Saders}, \& {Weiss}}]{Berger2020b}
{Berger}, T.~A., {Huber}, D., {Gaidos}, E., {van Saders}, J.~L., \& {Weiss},
  L.~M. 2020{\natexlab{a}}, \aj, 160, 108

\bibitem[{{Berger} {et~al.}(2020{\natexlab{b}}){Berger}, {Huber}, {van Saders},
  {Gaidos}, {Tayar}, \& {Kraus}}]{Berger2020a}
{Berger}, T.~A., {Huber}, D., {van Saders}, J.~L., {et~al.} 2020{\natexlab{b}},
  \aj, 159, 280

\bibitem[{{Borucki}(2017)}]{Borucki2017}
{Borucki}, W.~J. 2017, Proceedings of the American Philosophical Society, 161,
  38

\bibitem[{{Borucki} {et~al.}(2010){Borucki}, {Koch}, {Basri}, {Batalha},
  {Brown}, {Caldwell}, {Caldwell}, {Christensen-Dalsgaard}, {Cochran},
  {DeVore}, {Dunham}, {Dupree}, {Gautier}, {Geary}, {Gilliland}, {Gould},
  {Howell}, {Jenkins}, {Kondo}, {Latham}, {Marcy}, {Meibom}, {Kjeldsen},
  {Lissauer}, {Monet}, {Morrison}, {Sasselov}, {Tarter}, {Boss}, {Brownlee},
  {Owen}, {Buzasi}, {Charbonneau}, {Doyle}, {Fortney}, {Ford}, {Holman},
  {Seager}, {Steffen}, {Welsh}, {Rowe}, {Anderson}, {Buchhave}, {Ciardi},
  {Walkowicz}, {Sherry}, {Horch}, {Isaacson}, {Everett}, {Fischer}, {Torres},
  {Johnson}, {Endl}, {MacQueen}, {Bryson}, {Dotson}, {Haas}, {Kolodziejczak},
  {Van Cleve}, {Chandrasekaran}, {Twicken}, {Quintana}, {Clarke}, {Allen},
  {Li}, {Wu}, {Tenenbaum}, {Verner}, {Bruhweiler}, {Barnes}, \&
  {Prsa}}]{Borucki2010}
{Borucki}, W.~J., {Koch}, D., {Basri}, G., {et~al.} 2010, Science, 327, 977

\bibitem[{{Bryson} {et~al.}(2020{\natexlab{a}}){Bryson}, {Coughlin}, {Batalha},
  {Berger}, {Huber}, {Burke}, {Dotson}, \&
  {Mullally}}]{Bryson2020-OG-Reliability}
{Bryson}, S., {Coughlin}, J., {Batalha}, N.~M., {et~al.} 2020{\natexlab{a}},
  \aj, 159, 279

\bibitem[{{Bryson} {et~al.}(2020{\natexlab{b}}){Bryson}, {Coughlin},
  {Kunimoto}, \& {Mullally}}]{Bryson2020}
{Bryson}, S., {Coughlin}, J.~L., {Kunimoto}, M., \& {Mullally}, S.~E.
  2020{\natexlab{b}}, \aj, 160, 200

\bibitem[{{Bryson} {et~al.}(2021){Bryson}, {Kunimoto}, {Kopparapu}, {Coughlin},
  {Borucki}, {Koch}, {Aguirre}, {Allen}, {Barentsen}, {Batalha}, {Berger},
  {Boss}, {Buchhave}, {Burke}, {Caldwell}, {Campbell}, {Catanzarite},
  {Chandrasekaran}, {Chaplin}, {Christiansen}, {Christensen-Dalsgaard},
  {Ciardi}, {Clarke}, {Cochran}, {Dotson}, {Doyle}, {Duarte}, {Dunham},
  {Dupree}, {Endl}, {Fanson}, {Ford}, {Fujieh}, {Gautier}, {Geary},
  {Gilliland}, {Girouard}, {Gould}, {Haas}, {Henze}, {Holman}, {Howard},
  {Howell}, {Huber}, {Hunter}, {Jenkins}, {Kjeldsen}, {Kolodziejczak},
  {Larson}, {Latham}, {Li}, {Mathur}, {Meibom}, {Middour}, {Morris}, {Morton},
  {Mullally}, {Mullally}, {Pletcher}, {Prsa}, {Quinn}, {Quintana}, {Ragozzine},
  {Ramirez}, {Sanderfer}, {Sasselov}, {Seader}, {Shabram}, {Shporer}, {Smith},
  {Steffen}, {Still}, {Torres}, {Troeltzsch}, {Twicken}, {Uddin}, {Van Cleve},
  {Voss}, {Weiss}, {Welsh}, {Wohler}, \& {Zamudio}}]{Bryson2021}
{Bryson}, S., {Kunimoto}, M., {Kopparapu}, R.~K., {et~al.} 2021, \aj, 161, 36

\bibitem[{{Burke} \& {Catanzarite}(2017)}]{BurkeCatanzarite2017}
{Burke}, C.~J., \& {Catanzarite}, J. 2017, {Planet Detection Metrics:
  Per-Target Detection Contours for Data Release 25}, Kepler Science Document
  KSCI-19111-002, id. 19. Edited by Michael R. Haas and Natalie M. Batalha

\bibitem[{{Burke} {et~al.}(2015){Burke}, {Christiansen}, {Mullally}, {Seader},
  {Huber}, {Rowe}, {Coughlin}, {Thompson}, {Catanzarite}, {Clarke}, {Morton},
  {Caldwell}, {Bryson}, {Haas}, {Batalha}, {Jenkins}, {Tenenbaum}, {Twicken},
  {Li}, {Quintana}, {Barclay}, {Henze}, {Borucki}, {Howell}, \&
  {Still}}]{Burke2015}
{Burke}, C.~J., {Christiansen}, J.~L., {Mullally}, F., {et~al.} 2015, \apj,
  809, 8

\bibitem[{{Coughlin}(2017)}]{Coughlin2017}
{Coughlin}, J.~L. 2017, {Planet Detection Metrics: Robovetter Completeness and
  Effectiveness for Data Release 25}, Kepler Science Document KSCI-19114-002

\bibitem[{{Fernandes} {et~al.}(2019){Fernandes}, {Mulders}, {Pascucci},
  {Mordasini}, \& {Emsenhuber}}]{Fernandes2019}
{Fernandes}, R.~B., {Mulders}, G.~D., {Pascucci}, I., {Mordasini}, C., \&
  {Emsenhuber}, A. 2019, \apj, 874, 81

\bibitem[{Foreman-Mackey(2016)}]{corner}
Foreman-Mackey, D. 2016, The Journal of Open Source Software, 24,
  doi:10.21105/joss.00024

\bibitem[{{Foreman-Mackey} {et~al.}(2013){Foreman-Mackey}, {Hogg}, {Lang}, \&
  {Goodman}}]{ForemanMackey2012}
{Foreman-Mackey}, D., {Hogg}, D.~W., {Lang}, D., \& {Goodman}, J. 2013, \pasp,
  125, 306

\bibitem[{{Fulton} \& {Petigura}(2018)}]{FultonPetigura2018}
{Fulton}, B.~J., \& {Petigura}, E.~A. 2018, \aj, 156, 264

\bibitem[{{Fulton} {et~al.}(2017){Fulton}, {Petigura}, {Howard}, {Isaacson},
  {Marcy}, {Cargile}, {Hebb}, {Weiss}, {Johnson}, {Morton}, {Sinukoff},
  {Crossfield}, \& {Hirsch}}]{Fulton2017}
{Fulton}, B.~J., {Petigura}, E.~A., {Howard}, A.~W., {et~al.} 2017, \aj, 154,
  109

\bibitem[{{Gardner} {et~al.}(2006){Gardner}, {Mather}, {Clampin}, {Doyon},
  {Greenhouse}, {Hammel}, {Hutchings}, {Jakobsen}, {Lilly}, {Long}, {Lunine},
  {McCaughrean}, {Mountain}, {Nella}, {Rieke}, {Rieke}, {Rix}, {Smith},
  {Sonneborn}, {Stiavelli}, {Stockman}, {Windhorst}, \& {Wright}}]{JWST}
{Gardner}, J.~P., {Mather}, J.~C., {Clampin}, M., {et~al.} 2006, \ssr, 123, 485

\bibitem[{{Garrett} {et~al.}(2018){Garrett}, {Savransky}, \&
  {Belikov}}]{Garrett2018}
{Garrett}, D., {Savransky}, D., \& {Belikov}, R. 2018, \pasp, 130, 114403

\bibitem[{{Gaudi} {et~al.}(2020){Gaudi}, {Seager}, {Mennesson}, {Kiessling},
  {Warfield}, {Cahoy}, {Clarke}, {Domagal-Goldman}, {Feinberg}, {Guyon},
  {Kasdin}, {Mawet}, {Plavchan}, {Robinson}, {Rogers}, {Scowen}, {Somerville},
  {Stapelfeldt}, {Stark}, {Stern}, {Turnbull}, {Amini}, {Kuan}, {Martin},
  {Morgan}, {Redding}, {Stahl}, {Webb}, {Alvarez-Salazar}, {Arnold}, {Arya},
  {Balasubramanian}, {Baysinger}, {Bell}, {Below}, {Benson}, {Blais}, {Booth},
  {Bourgeois}, {Bradford}, {Brewer}, {Brooks}, {Cady}, {Caldwell}, {Calvet},
  {Carr}, {Chan}, {Cormarkovic}, {Coste}, {Cox}, {Danner}, {Davis}, {Dewell},
  {Dorsett}, {Dunn}, {East}, {Effinger}, {Eng}, {Freebury}, {Garcia}, {Gaskin},
  {Greene}, {Hennessy}, {Hilgemann}, {Hood}, {Holota}, {Howe}, {Huang}, {Hull},
  {Hunt}, {Hurd}, {Johnson}, {Kissil}, {Knight}, {Kolenz}, {Kraus}, {Krist},
  {Li}, {Lisman}, {Mandic}, {Mann}, {Marchen}, {Marrese-Reading}, {McCready},
  {McGown}, {Missun}, {Miyaguchi}, {Moore}, {Nemati}, {Nikzad}, {Nissen},
  {Novicki}, {Perrine}, {Pineda}, {Polanco}, {Putnam}, {Qureshi}, {Richards},
  {Eldorado Riggs}, {Rodgers}, {Rud}, {Saini}, {Scalisi}, {Scharf}, {Schulz},
  {Serabyn}, {Sigrist}, {Sikkia}, {Singleton}, {Shaklan}, {Smith}, {Southerd},
  {Stahl}, {Steeves}, {Sturges}, {Sullivan}, {Tang}, {Taras}, {Tesch},
  {Therrell}, {Tseng}, {Valente}, {Van Buren}, {Villalvazo}, {Warwick}, {Webb},
  {Westerhoff}, {Wofford}, {Wu}, {Woo}, {Wood}, {Ziemer}, {Arney}, {Anderson},
  {Ma{\'\i}z-Apell{\'a}niz}, {Bartlett}, {Belikov}, {Bendek}, {Cenko},
  {Douglas}, {Dulz}, {Evans}, {Faramaz}, {Feng}, {Ferguson}, {Follette},
  {Ford}, {Garc{\'\i}a}, {Geha}, {Gelino}, {G{\"o}tberg}, {Hildebrandt}, {Hu},
  {Jahnke}, {Kennedy}, {Kreidberg}, {Isella}, {Lopez}, {Marchis}, {Macri},
  {Marley}, {Matzko}, {Mazoyer}, {McCandliss}, {Meshkat}, {Mordasini},
  {Morris}, {Nielsen}, {Newman}, {Petigura}, {Postman}, {Reines}, {Roberge},
  {Roederer}, {Ruane}, {Schwieterman}, {Sirbu}, {Spalding}, {Teplitz},
  {Tumlinson}, {Turner}, {Werk}, {Wofford}, {Wyatt}, {Young}, \&
  {Zellem}}]{HabEx}
{Gaudi}, B.~S., {Seager}, S., {Mennesson}, B., {et~al.} 2020, arXiv e-prints,
  arXiv:2001.06683

\bibitem[{{Ginzburg} {et~al.}(2016){Ginzburg}, {Schlichting}, \&
  {Sari}}]{Ginzburg2016}
{Ginzburg}, S., {Schlichting}, H.~E., \& {Sari}, R. 2016, \apj, 825, 29

\bibitem[{{Ginzburg} {et~al.}(2018){Ginzburg}, {Schlichting}, \&
  {Sari}}]{Ginzburg2018}
---. 2018, \mnras, 476, 759

\bibitem[{{Gupta} \& {Schlichting}(2019)}]{GuptaSchlichting2019}
{Gupta}, A., \& {Schlichting}, H.~E. 2019, \mnras, 487, 24

\bibitem[{{Gupta} \& {Schlichting}(2020)}]{GuptaSchlichting2020}
---. 2020, \mnras, 493, 792

\bibitem[{{Hardegree-Ullman} {et~al.}(2020){Hardegree-Ullman}, {Zink},
  {Christiansen}, {Dressing}, {Ciardi}, \& {Schlieder}}]{KHU2020}
{Hardegree-Ullman}, K.~K., {Zink}, J.~K., {Christiansen}, J.~L., {et~al.} 2020,
  \apjs, 247, 28

\bibitem[{{Hart}(1978)}]{Hart1978}
{Hart}, M.~H. 1978, \icarus, 33, 23

\bibitem[{{He} {et~al.}(2021){He}, {Ford}, \& {Ragozzine}}]{He2021}
{He}, M.~Y., {Ford}, E.~B., \& {Ragozzine}, D. 2021, \aj, 161, 16

\bibitem[{{Howard} {et~al.}(2012){Howard}, {Marcy}, {Bryson}, {Jenkins},
  {Rowe}, {Batalha}, {Borucki}, {Koch}, {Dunham}, {Gautier}, {Van Cleve},
  {Cochran}, {Latham}, {Lissauer}, {Torres}, {Brown}, {Gilliland}, {Buchhave},
  {Caldwell}, {Christensen-Dalsgaard}, {Ciardi}, {Fressin}, {Haas}, {Howell},
  {Kjeldsen}, {Seager}, {Rogers}, {Sasselov}, {Steffen}, {Basri},
  {Charbonneau}, {Christiansen}, {Clarke}, {Dupree}, {Fabrycky}, {Fischer},
  {Ford}, {Fortney}, {Tarter}, {Girouard}, {Holman}, {Johnson}, {Klaus},
  {Machalek}, {Moorhead}, {Morehead}, {Ragozzine}, {Tenenbaum}, {Twicken},
  {Quinn}, {Isaacson}, {Shporer}, {Lucas}, {Walkowicz}, {Welsh}, {Boss},
  {Devore}, {Gould}, {Smith}, {Morris}, {Prsa}, {Morton}, {Still}, {Thompson},
  {Mullally}, {Endl}, \& {MacQueen}}]{Howard2012}
{Howard}, A.~W., {Marcy}, G.~W., {Bryson}, S.~T., {et~al.} 2012, \apjs, 201, 15

\bibitem[{{Hsu} {et~al.}(2019){Hsu}, {Ford}, {Ragozzine}, \& {Ashby}}]{Hsu2019}
{Hsu}, D.~C., {Ford}, E.~B., {Ragozzine}, D., \& {Ashby}, K. 2019, \aj, 158,
  109

\bibitem[{{Huang}(1959)}]{Huang1959}
{Huang}, S.-S. 1959, American Scientist, 47, 397

\bibitem[{Hunter(2007)}]{pyplot}
Hunter, J.~D. 2007, Computing in Science {\&} Engineering, 9, 90

\bibitem[{{Jin} {et~al.}(2014){Jin}, {Mordasini}, {Parmentier}, {van Boekel},
  {Henning}, \& {Ji}}]{Jin2014}
{Jin}, S., {Mordasini}, C., {Parmentier}, V., {et~al.} 2014, \apj, 795, 65

\bibitem[{{Johnstone} {et~al.}(2021){Johnstone}, {Bartel}, \&
  {G{\"u}del}}]{Johnstone2021}
{Johnstone}, C.~P., {Bartel}, M., \& {G{\"u}del}, M. 2021, \aap, 649, A96

\bibitem[{Jones {et~al.}(2001--)Jones, Oliphant, Peterson, {et~al.}}]{scipy}
Jones, E., Oliphant, T., Peterson, P., {et~al.} 2001--, {SciPy}: Open source
  scientific tools for {Python}, [Online; accessed <today>]

\bibitem[{{Kite} {et~al.}(2019){Kite}, {Fegley}, {Schaefer}, \&
  {Ford}}]{Kite2019}
{Kite}, E.~S., {Fegley}, Bruce, J., {Schaefer}, L., \& {Ford}, E.~B. 2019,
  \apjl, 887, L33

\bibitem[{{Kopparapu} {et~al.}(2013){Kopparapu}, {Ramirez}, {Kasting}, {Eymet},
  {Robinson}, {Mahadevan}, {Terrien}, {Domagal-Goldman}, {Meadows}, \&
  {Deshpande}}]{Kopparapu2013}
{Kopparapu}, R.~K., {Ramirez}, R., {Kasting}, J.~F., {et~al.} 2013, \apj, 765,
  131

\bibitem[{{Kopparapu} {et~al.}(2018){Kopparapu}, {H{\'e}brard}, {Belikov},
  {Batalha}, {Mulders}, {Stark}, {Teal}, {Domagal-Goldman}, \&
  {Mandell}}]{Kopparapu2018}
{Kopparapu}, R.~K., {H{\'e}brard}, E., {Belikov}, R., {et~al.} 2018, \apj, 856,
  122

\bibitem[{{Kunimoto} \& {Matthews}(2020)}]{Kunimoto&Matthews2020}
{Kunimoto}, M., \& {Matthews}, J.~M. 2020, \aj, 159, 248

\bibitem[{{Lammer} {et~al.}(2003){Lammer}, {Selsis}, {Ribas}, {Guinan},
  {Bauer}, \& {Weiss}}]{Lammer2003}
{Lammer}, H., {Selsis}, F., {Ribas}, I., {et~al.} 2003, \apjl, 598, L121

\bibitem[{{Lee} {et~al.}(2022){Lee}, {Karalis}, \& {Thorngren}}]{Lee2022}
{Lee}, E.~J., {Karalis}, A., \& {Thorngren}, D.~P. 2022, arXiv e-prints,
  arXiv:2201.09898

\bibitem[{{Lopez} \& {Fortney}(2013)}]{Lopez&Fortney2013}
{Lopez}, E.~D., \& {Fortney}, J.~J. 2013, \apj, 776, 2

\bibitem[{{Lopez} \& {Fortney}(2014)}]{Lopez&Fortney2014}
---. 2014, \apj, 792, 1

\bibitem[{{Lopez} \& {Rice}(2018)}]{Lopez&Rice2018}
{Lopez}, E.~D., \& {Rice}, K. 2018, \mnras, 479, 5303

\bibitem[{{Loyd} {et~al.}(2020){Loyd}, {Shkolnik}, {Schneider},
  {Richey-Yowell}, {Barman}, {Peacock}, \& {Pagano}}]{Loyd2020}
{Loyd}, R.~O.~P., {Shkolnik}, E.~L., {Schneider}, A.~C., {et~al.} 2020, \apj,
  890, 23

\bibitem[{{Martinez} {et~al.}(2019){Martinez}, {Cunha}, {Ghezzi}, \&
  {Smith}}]{Martinez2019}
{Martinez}, C.~F., {Cunha}, K., {Ghezzi}, L., \& {Smith}, V.~V. 2019, \apj,
  875, 29

\bibitem[{{Moe} \& {Kratter}(2021)}]{Moe&Kratter2021}
{Moe}, M., \& {Kratter}, K.~M. 2021, \mnras, 507, 3593

\bibitem[{{Morton} {et~al.}(2016){Morton}, {Bryson}, {Coughlin}, {Rowe},
  {Ravichandran}, {Petigura}, {Haas}, \& {Batalha}}]{Morton2016}
{Morton}, T.~D., {Bryson}, S.~T., {Coughlin}, J.~L., {et~al.} 2016, \apj, 822,
  86

\bibitem[{{Mulders} {et~al.}(2019){Mulders}, {Mordasini}, {Pascucci}, {Ciesla},
  {Emsenhuber}, \& {Apai}}]{Mulders2019}
{Mulders}, G.~D., {Mordasini}, C., {Pascucci}, I., {et~al.} 2019, \apj, 887,
  157

\bibitem[{{Mulders} {et~al.}(2020){Mulders}, {O'Brien}, {Ciesla}, {Apai}, \&
  {Pascucci}}]{Mulders2020}
{Mulders}, G.~D., {O'Brien}, D.~P., {Ciesla}, F.~J., {Apai}, D., \& {Pascucci},
  I. 2020, \apj, 897, 72

\bibitem[{{Mulders} {et~al.}(2015){Mulders}, {Pascucci}, \&
  {Apai}}]{Mulders2015}
{Mulders}, G.~D., {Pascucci}, I., \& {Apai}, D. 2015, \apj, 798, 112

\bibitem[{{Mulders} {et~al.}(2018){Mulders}, {Pascucci}, {Apai}, \&
  {Ciesla}}]{Mulders2018}
{Mulders}, G.~D., {Pascucci}, I., {Apai}, D., \& {Ciesla}, F.~J. 2018, \aj,
  156, 24

\bibitem[{{National Academies of Sciences, Engineering, and
  Medicine}(2021)}]{Decadal}
{National Academies of Sciences, Engineering, and Medicine}. 2021, Pathways to
  Discovery in Astronomy and Astrophysics for the 2020s (Washington, DC: The
  National Academies Press), doi:10.17226/26141

\bibitem[{{Neil} \& {Rogers}(2020)}]{Neil2020}
{Neil}, A.~R., \& {Rogers}, L.~A. 2020, \apj, 891, 12

\bibitem[{{Owen} \& {Wu}(2013)}]{Owen&Wu2013}
{Owen}, J.~E., \& {Wu}, Y. 2013, \apj, 775, 105

\bibitem[{{Owen} \& {Wu}(2017)}]{Owen&Wu2017}
---. 2017, \apj, 847, 29

\bibitem[{{Pascucci} {et~al.}(2018){Pascucci}, {Mulders}, {Gould}, \&
  {Fernandes}}]{Pascucci2018}
{Pascucci}, I., {Mulders}, G.~D., {Gould}, A., \& {Fernandes}, R. 2018, \apjl,
  856, L28

\bibitem[{{Pascucci} {et~al.}(2019){Pascucci}, {Mulders}, \&
  {Lopez}}]{Pascucci2019}
{Pascucci}, I., {Mulders}, G.~D., \& {Lopez}, E. 2019, \apjl, 883, L15

\bibitem[{{Pascucci} {et~al.}(2016){Pascucci}, {Testi}, {Herczeg}, {Long},
  {Manara}, {Hendler}, {Mulders}, {Krijt}, {Ciesla}, {Henning}, {Mohanty},
  {Drabek-Maunder}, {Apai}, {Sz{\H{u}}cs}, {Sacco}, \&
  {Olofsson}}]{Pascucci2016}
{Pascucci}, I., {Testi}, L., {Herczeg}, G.~J., {et~al.} 2016, \apj, 831, 125

\bibitem[{{Pecaut} \& {Mamajek}(2013)}]{MamajekTable}
{Pecaut}, M.~J., \& {Mamajek}, E.~E. 2013, \apjs, 208, 9

\bibitem[{{Petigura} {et~al.}(2013{\natexlab{a}}){Petigura}, {Howard}, \&
  {Marcy}}]{Petigura2013}
{Petigura}, E.~A., {Howard}, A.~W., \& {Marcy}, G.~W. 2013{\natexlab{a}},
  Proceedings of the National Academy of Science, 110, 19273

\bibitem[{{Petigura} {et~al.}(2013{\natexlab{b}}){Petigura}, {Marcy}, \&
  {Howard}}]{Petigura2013_Plateau}
{Petigura}, E.~A., {Marcy}, G.~W., \& {Howard}, A.~W. 2013{\natexlab{b}}, \apj,
  770, 69

\bibitem[{{Petigura} {et~al.}(2018){Petigura}, {Marcy}, {Winn}, {Weiss},
  {Fulton}, {Howard}, {Sinukoff}, {Isaacson}, {Morton}, \&
  {Johnson}}]{Petigura2018}
{Petigura}, E.~A., {Marcy}, G.~W., {Winn}, J.~N., {et~al.} 2018, \aj, 155, 89

\bibitem[{{Rogers} {et~al.}(2021){Rogers}, {Gupta}, {Owen}, \&
  {Schlichting}}]{Rogers2021}
{Rogers}, J.~G., {Gupta}, A., {Owen}, J.~E., \& {Schlichting}, H.~E. 2021,
  \mnras, 508, 5886

\bibitem[{{Rogers} \& {Owen}(2021)}]{RogersOwen2021}
{Rogers}, J.~G., \& {Owen}, J.~E. 2021, \mnras, 503, 1526

\bibitem[{{Rogers}(2015)}]{Rogers2015}
{Rogers}, L.~A. 2015, \apj, 801, 41

\bibitem[{{Sandoval} {et~al.}(2021){Sandoval}, {Contardo}, \&
  {David}}]{Sandoval2021}
{Sandoval}, A., {Contardo}, G., \& {David}, T.~J. 2021, \apj, 911, 117

\bibitem[{{The LUVOIR Team}(2019)}]{LUVOIR}
{The LUVOIR Team}. 2019, arXiv e-prints, arXiv:1912.06219

\bibitem[{{Thompson} {et~al.}(2018){Thompson}, {Coughlin}, {Hoffman},
  {Mullally}, {Christiansen}, {Burke}, {Bryson}, {Batalha}, {Haas},
  {Catanzarite}, {Rowe}, {Barentsen}, {Caldwell}, {Clarke}, {Jenkins}, {Li},
  {Latham}, {Lissauer}, {Mathur}, {Morris}, {Seader}, {Smith}, {Klaus},
  {Twicken}, {Van Cleve}, {Wohler}, {Akeson}, {Ciardi}, {Cochran}, {Henze},
  {Howell}, {Huber}, {Pr{\v{s}}a}, {Ram{\'\i}rez}, {Morton}, {Barclay},
  {Campbell}, {Chaplin}, {Charbonneau}, {Christensen-Dalsgaard}, {Dotson},
  {Doyle}, {Dunham}, {Dupree}, {Ford}, {Geary}, {Girouard}, {Isaacson},
  {Kjeldsen}, {Quintana}, {Ragozzine}, {Shabram}, {Shporer}, {Silva Aguirre},
  {Steffen}, {Still}, {Tenenbaum}, {Welsh}, {Wolfgang}, {Zamudio}, {Koch}, \&
  {Borucki}}]{Thompson2018}
{Thompson}, S.~E., {Coughlin}, J.~L., {Hoffman}, K., {et~al.} 2018, \apjs, 235,
  38

\bibitem[{van~der Walt {et~al.}(2011)van~der Walt, Colbert, \&
  Varoquaux}]{numpy}
van~der Walt, S., Colbert, S.~C., \& Varoquaux, G. 2011, Computing in Science
  {\&} Engineering, 13, 22

\bibitem[{{Van Eylen} {et~al.}(2018){Van Eylen}, {Agentoft}, {Lundkvist},
  {Kjeldsen}, {Owen}, {Fulton}, {Petigura}, \& {Snellen}}]{VanEylen2018}
{Van Eylen}, V., {Agentoft}, C., {Lundkvist}, M.~S., {et~al.} 2018, \mnras,
  479, 4786

\bibitem[{{Wu}(2019)}]{Wu2019}
{Wu}, Y. 2019, \apj, 874, 91

\bibitem[{{Yang} {et~al.}(2020){Yang}, {Xie}, \& {Zhou}}]{Yang2020}
{Yang}, J.-Y., {Xie}, J.-W., \& {Zhou}, J.-L. 2020, \aj, 159, 164

\bibitem[{{Youdin}(2011)}]{Youdin2011}
{Youdin}, A.~N. 2011, \apj, 742, 38

\bibitem[{{Zahnle} \& {Catling}(2017)}]{ZahnleCatling2017}
{Zahnle}, K.~J., \& {Catling}, D.~C. 2017, \apj, 843, 122

\bibitem[{{Zhu} \& {Dong}(2021)}]{ZhuDong2021}
{Zhu}, W., \& {Dong}, S. 2021, \araa, 59, 291

\bibitem[{{Zink} {et~al.}(2019){Zink}, {Christiansen}, \& {Hansen}}]{Zink2019}
{Zink}, J.~K., {Christiansen}, J.~L., \& {Hansen}, B. M.~S. 2019, \mnras, 483,
  4479

\end{thebibliography}
\clearpage

\appendix
\section{Scaling the Radius Valley by Stellar Mass}\label{sec:Appendix_RVscaling}

\begin{figure*}
    \centering
    \includegraphics[width = 0.9\textwidth]{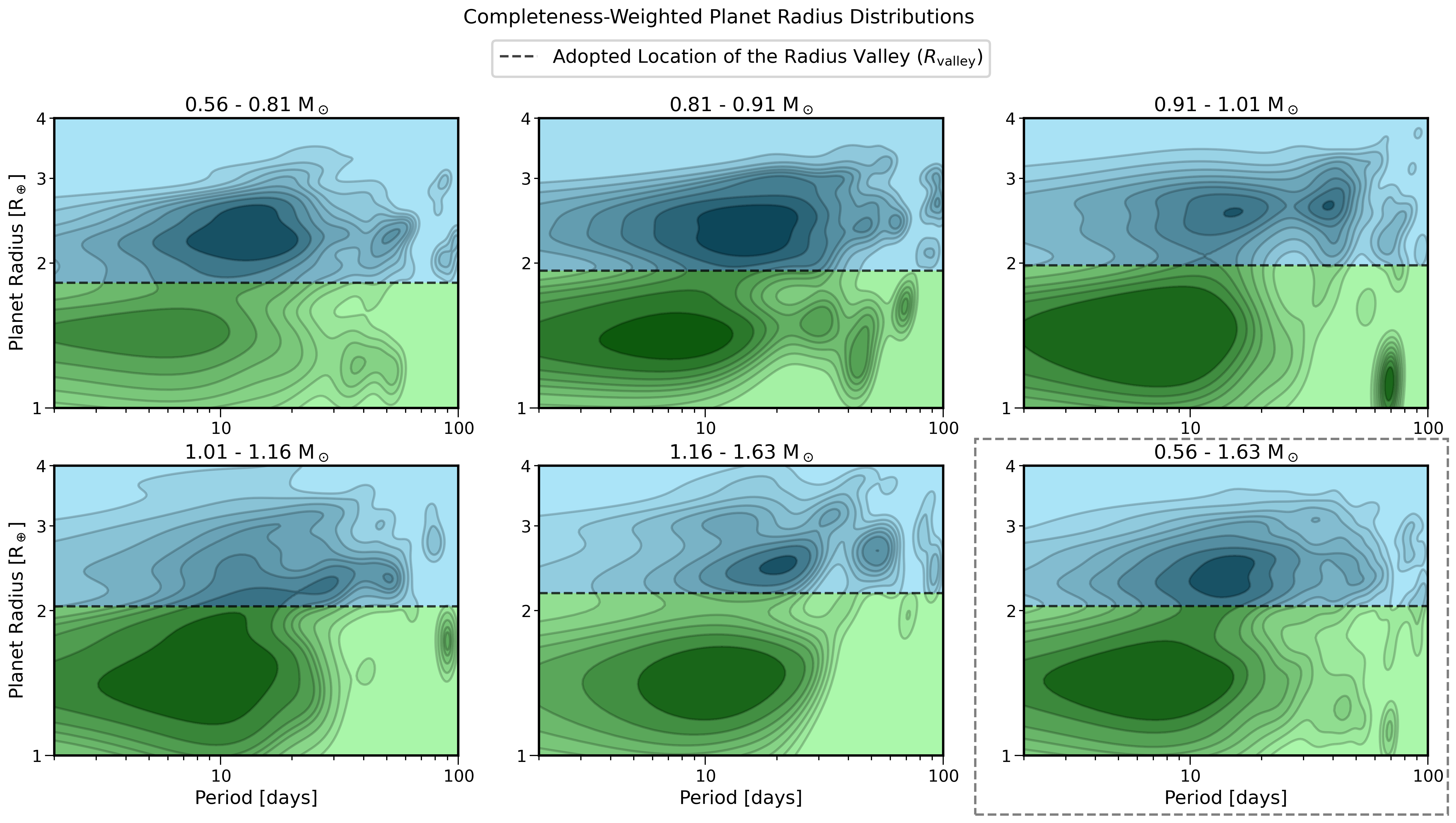}
    \caption{Orbital period and radius distributions for the planets in our samples, with density contours weighted by planet completeness. Each panel represents a different stellar mass bin, with the full FGK sample shown in the lower-right. The location of the radius valley (black dashed line) follows the scaling of \citet{Wu2019}, and is used to separate the super-Earth (green) and sub-Neptune (blue) populations.}
    \label{fig:PRContour}
\end{figure*}

This work adopts a scaling of the radius valley which depends on the average stellar mass in each bin, of the form $R_\mathrm{p, adj.} = R_\mathrm{p} (\frac{\Mstar}{\SolarMass})^{1/4}$ following the suggestion of \citet{Wu2019}. Figure~(\ref{fig:PRContour}) shows two-dimensional density contour plots based on the completeness-weighted planet orbital period and radius distributions in each stellar mass bin. With the exception of the $\left[1.01, 1.16\right]\,\SolarMass$ bin, each panel shows the clear presence of a low-density region between super-Earth and sub-Neptune peaks, which we interpret as the radius valley. The scaling relation's normalization of $R_\mathrm{valley} = 2\,\EarthRad$ at $\Mstar = 1\,\SolarMass$ is chosen by visual inspection to comply with the observed low-density regions of Figure~(\ref{fig:PRContour}).

While the $\left[1.01, 1.16\right]\,\SolarMass$ bin lacks such a pronounced valley, the fractional occurrence curves in this bin still follow the same functional form of other bins in this work (see Figure~\ref{fig:FracOcc}). As there is no compelling evidence suggesting a lack of radius valley around this particular sample of stars, we adopt the same normalization for all bins. The stellar mass-scaled values of $R_\mathrm{valley}$ for each bin are presented in Table (\ref{tab:MassBins}), and are considered as the boundary between the super-Earth and sub-Neptune populations.

\citet{Lopez&Rice2018} suggest that, if the remnant cores of sub-Neptunes comprise most of the bare rocky planet population, then the radius valley (or ``transition radius" between super-Earths and sub-Neptunes) should decrease with increasing orbital period. Additionally, both models of photoevaporation and core-powered mass loss predict a relatively flat, slightly negative slope in the log orbital period - log radius plane: $\dd \ln{R_\mathrm{valley}} / \dd \ln{P} = -0.16$ or $-0.11$, respectively \citep{Rogers2021}. These predictions have been validated by observational works which also recover the weak trend with orbital period \citep{VanEylen2018, Martinez2019}. There are also suggestions of different forms for the radius valley's dependence on stellar mass \citep{Rogers2015}, but degeneracy with planet core mass make their validity difficult to discern. In terms of both orbital period and stellar mass, we opt for the relation of \citet{Wu2019} which matches our data well. Incorporating a radius valley that scales with orbital period would significantly alter the functional form, dependencies, and normalization of our occurrence distribution model. Thus, testing various forms of the radius valley would be a compelling exercise beyond the scope of this work. 

\section{Updates to the EPOS package.}\label{epos4.0}

In this section, we outline the new additions and modifications to \epos{}, the Exoplanet Population Observation Simulator \citep{Mulders2018}. These changes are only relevant to the \kepler transit survey, and do not apply to \epos{}'s handling of radial velocity survey data. We have updated the catalog of stellar properties within \epos{} to those provided by the \gaia-\kepler Stellar Properties Catalog \citep{Berger2020a}. Previously, \epos{} users could select a generic subset of FGKM dwarf stars using a predetermined set of internal classifications. Because definitions of spectral type differ across the community, users can now specify their own samples based on custom stellar mass ranges. The \epos{} package has been updated to automatically calculate the corresponding detection and vetting efficiencies, which are then stored locally. Detailed descriptions of these calculations can be found in Section 2.3 and Appendix B of \citet{Mulders2018}.

Recent works have shown that different \kepler catalogs produce statistically inconsistent occurrence rates unless they are corrected by both completeness and reliability \citep{Bryson2020}. Because of this, we have added an option within \epos{} to include vetting reliability in planet occurrence calculations. Normally, \epos{} calculates occurrence rates using the inverse detection efficiency method. For each observed planet, the survey completeness (including the vetting efficiency) is evaluated at that planet's radius and orbital period. If the reliability argument is supplied, the per-planet occurrence rate will then be multiplied by the per-planet vetting reliability. The updated version of \epos{} contains a file of the per-planet vetting reliability for the \kepler sample, calculated following the methods and code of \citet{Bryson2020}. As mentioned in Section~(\ref{sec:epos}), vetting reliability is intended to circumvent the typical disposition score cuts used to isolate highly reliable planets. Thus by accounting for vetting reliability, the vetting efficiency component is not necessary, and so we do not limit our sample by disposition score in this work. 


\section{Description of the Parametric Model.}\label{FitDetails}
Here we detail the process used to fit a parametric model to the small \kepler exoplanet population, including descriptions of the functional form and fitting procedure. \citet{Youdin2011} outlines how a planet population can be described through a planet distribution function (PLDF) by way of a parametric model which depends on planet properties. This approach is adapted by \citet{Burke2015} to allow for more complicated or higher dimensional models, and is used to fit a PLDF to a sample of \kepler planets ($0.75 \leq \Rp \leq 2.5\,\EarthRad, 50 \leq \Porb \leq 300$\,days) around GK dwarf stars. In both works, the distribution function is tailored to an observed data set through likelihood estimation techniques, which we describe in Section (\ref{sec:MLE}). 

\subsection{The Planet Distribution Function} \label{sec:pldf}

As mentioned in Section (\ref{sec:model}), the general form of a PLDF used to calculate planet occurrence can be written as
\begin{equation} 
    \frac{\mathrm{d}^2 f}{\mathrm{d} \Porb \mathrm{d} \Rp} = F_\mathrm{0} C_\mathrm{n} g(\Porb,\Rp).
\end{equation}
$F_\mathrm{0}$ can be treated as the average number of planets per star, while $C_\mathrm{n}$ is a normalization factor with the requirement: 
\begin{equation}
    \int\limits_{P_{\mathrm{min}}}^{P_{\mathrm{max}}} \int\limits_{R_{\mathrm{min}}}^{R_{\mathrm{max}}} C_\mathrm{n} g(\Porb,\Rp) \mathrm{d} \Porb \mathrm{d} \Rp = 1.
\end{equation}
The PLDF is specifically defined (and fit) over the domain $P_{\mathrm{min}} \leq \Porb \leq P_{\mathrm{max}}$ and $R_{\mathrm{min}} \leq \Rp \leq R_{\mathrm{max}}$. In our case, the domain is $2 < \Porb < 100$\,days and $1 \leq \Rp \leq 3.5\,\EarthRad$ as chosen in Section~(\ref{sec:samples}).

The greatest appeal for this definition of the PLDF lies in the fully customizable description of the distribution's behavior, the shape function $g(\textbf{x})$. In the case of \citet{Burke2015}, the shape function consists of a power law in $\Porb$ and a broken power law in $\Rp$. In this work, the shape function is the product of two components: a broken power law in orbital period, $g_1(\Porb)$, which governs the overall distribution of small planets; and an orbital period-dependent function for the fractional occurrence of either super-Earths or sub-Neptunes, $g_2(\Porb, \Rp)$. Therefore, the distribution is governed by:
\begin{equation}
    g(\Porb,\Rp) = g_1(\Porb) \cdot g_2(\Porb,\Rp),
\end{equation}
where the occurrence at a given orbital period provided by $g_1$ is further distributed among the super-Earth and sub-Neptune bins by their orbital period-dependent fractions in $g_2$. The overarching small planet distribution is written as:
\begin{equation}
    g_1(\Porb) = 
    \begin{cases}
    (\Porb / P_\mathrm{break})^{\beta_1} & \text{if $P < P_\mathrm{break}$} \\
    (\Porb / P_\mathrm{break})^{\beta_2} & \text{if $P \geq P_\mathrm{break}$,}
    \end{cases}
\end{equation}
where the two $\Porb$ power law exponents ($\beta_1$ and $\beta_2$) govern the distribution on either side of the orbital period break $P_\mathrm{break}$.

\begin{figure*}
    \centering
    \includegraphics[scale=0.4]{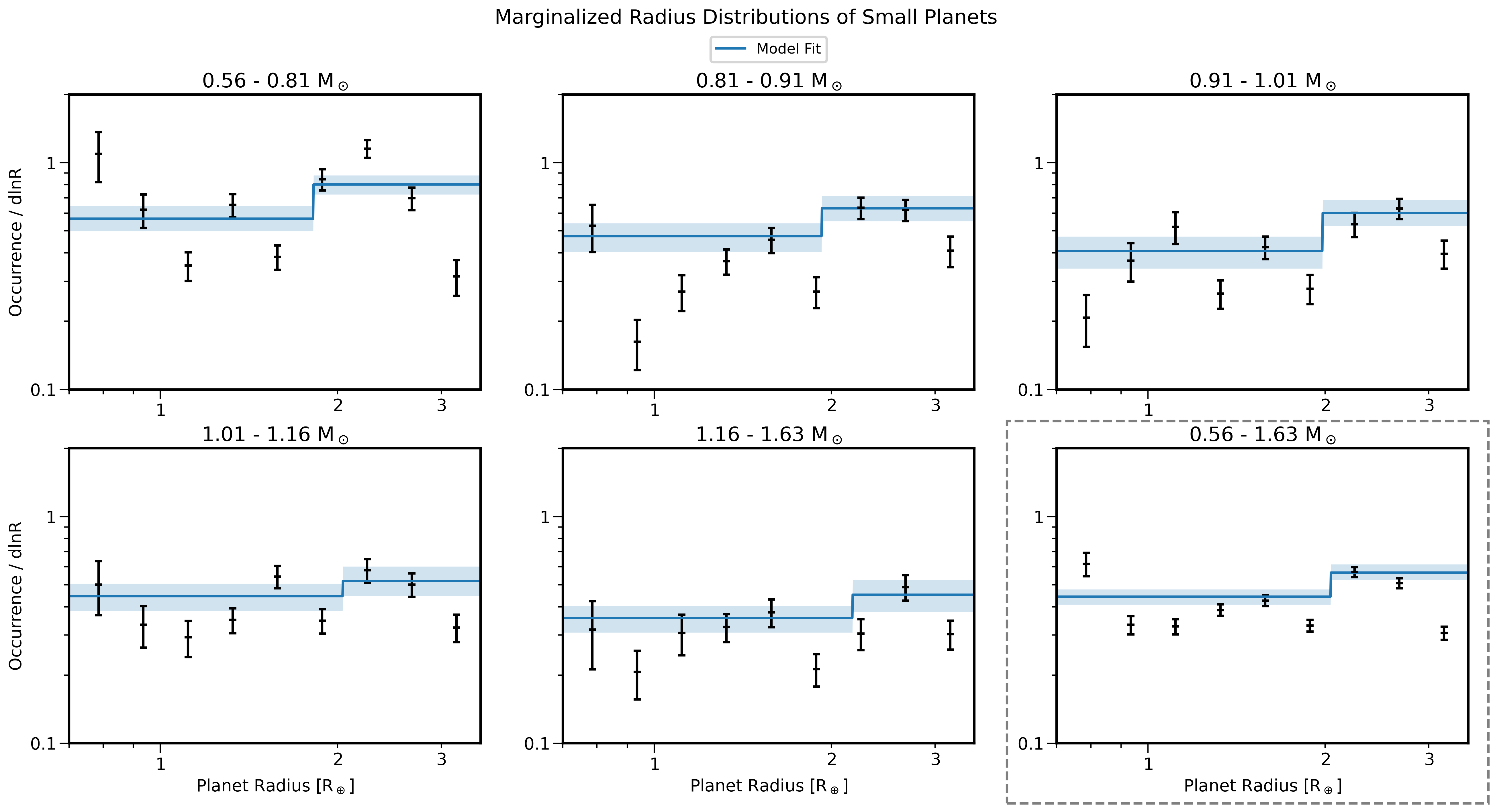}
    \caption{The occurrence of small planets marginalized over orbital period ($2 \leq \Porb \leq 100$\,days) as a function of planet radius. Each panel represents a different stellar mass bin, with the full FGK sample shown in the lower-right. Black points are the observed marginalized occurrence as calculated by \epos{} for a selection of planet radius bins. Blue lines indicate the best fit distributions evaluated for that bin.}
    \label{fig:MarginalizedRadiusDist}
\end{figure*}

Unlike \citet{Burke2015}, we do not see strong evidence for features more complex than a constant occurrence distribution in $\log{\Rp}$ when evaluating the marginalized radius distributions with \epos{} (see Figure~\ref{fig:MarginalizedRadiusDist}, similar to Figure 5 in \citealp{FultonPetigura2018}). This ``plateau" \citep{Petigura2013_Plateau} in the size distribution of small planets is often seen with the inverse detection efficiency method as used in this work and in \citet{FultonPetigura2018}. However, this method may underestimate the occurrence of sub-Earth planets in the size regime where sensitivity is low \citep{ZhuDong2021}. Other methods (e.g., the approximate Bayesian approach of \citealp{Hsu2019}) may be more accurate in their predictions of increasing occurrence at smaller sizes, as discussed in \citet{Lee2022}. As there is not yet a clear consensus for the functional form of this complex size distribution for small close-in planets, we do not adopt any functional dependence on the planet radius outside of what is described below.

Fractional occurrence is a metric which involves first computing the occurrence of super-Earths and sub-Neptunes, then dividing either by the sum of both. Yet the approach outlined above computes the occurrence (not fractional occurrence). Thus, we require a method of incorporating some behavior into the shape function such that, when fractional occurrence is computed, the desired trend is recovered. We first define some function $G(\Porb, \Rp)$ to describe the trend in fractional occurrence, and then reverse the calculation of fractional occurrence to discern the form of $g_2(\Porb, \Rp)$, the shape function component applied to normal occurrence which recovers the observed trend in fractional occurrence.

For the function $G(\Porb, \Rp)$, we opt to describe the behavior seen in Figure~(\ref{fig:FracOcc}) with four free parameters. The short-period regime is constrained by the asymptotic fraction $\chi_1$, which is fit simultaneously as $\chi_1$ for super-Earths and $\left(1-\chi_1\right)$ for sub-Neptunes. Similarly, the long-period population is fit by a separate fraction $\chi_2$ for super-Earths and $\left(1-\chi_2\right)$ for sub-Neptunes. These two regimes are separated by an orbital period $P_\mathrm{central}$ central to the adopted curvature, which has a smoothness $s$ constraining how quickly the curve flattens off towards $\chi_1$ or $\chi_2$ (or their complements, in the case of sub-Neptunes) moving away from $P_\mathrm{central}$. The curve itself is described by a hyperbolic tangent in $\log_{10}$ orbital period, which is normalized on $y =  \left[0,1\right]$ with plateaus scaled by $\chi_1$ for $\Porb \ll P_\mathrm{central}$ and $\chi_2$ for $\Porb \gg P_\mathrm{central}$. The function is thus:
\begin{equation}
    G(\Porb, \Rp) = \begin{cases}
    t \cdot \chi_1  + \left[1 - t\right] \cdot \chi_2 & \text{if $\Rp < R_\mathrm{valley}$} \\
    t \cdot \left(1-\chi_1\right)  + \left[1 - t\right] \cdot \left(1-\chi_2\right) & \text{if $\Rp > R_\mathrm{valley}$,} \\
    \end{cases}
\end{equation}
where
\begin{equation}
    t(\Porb) = 0.5 - 0.5 \tanh{\left(\frac{\log_{10}{\Porb} - \log_{10}{P_\mathrm{central}}}{\log_{10}{s}}\right)}.
\end{equation}
By these definitions, the super-Earth and sub-Neptune curves are always complementary. We note that $P_\mathrm{central}$ is not necessarily the intersection point $P_\mathrm{trans}$ where super-Earths and sub-Neptunes are present in equal abundance, except in the case where $\chi_1 = 1 - \chi_2$ such that the curve is vertically symmetric about the fraction $0.5$.

To implement this behavior into the PLDF via the shape function, we consider the definition of fractional occurrence as the occurrence of either super-Earths or sub-Neptunes divided by the occurrence of both. Since we do not adopt any planet radius dependence outside of binning super-Earths or sub-Neptunes, $G$ is only a function of orbital period when remaining within one of the two planet radius bins. We assume the same of $g_2$, confirmed by the final derivation. We also introduce a shorthand for bounds of integration where $a = R_{min}$, $b = R_{valley}$ and $c = R_{max}$. As an example case, the fractional occurrence of super-Earths is:
\begin{equation}
    G(P) = \frac{\int_a^b \frac{\dd f}{\dd P \dd R} \dd R}{\int_a^c \frac{\dd f}{\dd P \dd R} \dd R}.
\end{equation}
Because the normalization parameters and broken power law $g_1$ are independent of planet radius, they can be pulled from the integrals and divided out, such that:
\begin{equation}\label{eqn:G-integrate}
    G(P) = \frac{\int_a^b g_2(P) \dd R}{\int_a^c g_2(P) \dd R} = \frac{g_2(P) [R]_a^b}{g_2(P) [R]_a^c} = \frac{g_2(P) (b - a)}{g_2(P) (b-a) + (1-g_2(P))(c-b)}.
\end{equation}
In the denominator of the RHS, we note that the integral of $g_2$ is split into the super-Earth and sub-Neptune regimes, where the shape function must return the complement $\left(1-g_2\right)$ in the latter case. Solving for $g_2$, we find:
\begin{equation}\label{eqn:g2}
    g_2(P) = \frac{G(P) (c-b)}{(b-a) + G(P)(c-b) - G(P)(b-a)}.
\end{equation}
This is the form which is included in the shape function of our PLDF. Note that by their complementary nature, $\left(1 - g_2\right)$ recalls the sub-Neptune behavior.

As an additional caveat, the PLDF we use is defined in linear space for $\dd \Porb$ and $\dd R_p$. However the default output of \epos{} is binned in natural log space for $\dd \ln{\Porb}$ and $\dd \ln{R_p}$. We choose to fit directly to \kepler data in linear space, and then match the binned occurrence data in \epos{} after the fact, which is done with two modifications. We leave $g_1$, the broken power law in orbital period, as-is and append an extra factor of orbital period (analogous to adding $+1$ to the exponents $\beta_1$ and $\beta_2$) when plotting the fit results in log-orbital period. We approximate the marginalized radius distribution as constant in log-radius, which corresponds to an additional factor of $1/R$ in linear space appended to $g_2$. This suggests the integration in Equation (\ref{eqn:G-integrate}) is not independent of $R$, but the adjustment is a replacement of $R$ with $\ln{R}$ in the final product. Thus there is a modification which involves replacing $a$, $b$, and $c$ in Equation (\ref{eqn:g2}) with $\ln{a}$, $\ln{b}$, and $\ln{c}$.

In summary, the contribution of super-Earths to the combined pool of super-Earths and sub-Neptunes is expected to change as a function of orbital period, from $\chi_1$ ($P \ll P_\mathrm{central}$) to $\chi_2$ ($P \gg P_\mathrm{central}$). Here we have used ``$\ll$" and ``$\gg$" to indicate being far enough away from the central period $P_\mathrm{central}$ such that the fractional occurrence has sufficiently plateaued towards the asymptotic fractions and is not still dominated by curvature close to $P_\mathrm{central}$. The sub-Neptunes follow a similar fractional distribution of $\left(1-\chi_1\right)$, and $\left(1-\chi_2\right)$, respectively. Both super-Earths and sub-Neptunes are further subjected to the broken power law with  $g_1(\Porb)$ - we do not make any requirements on the relationship between $P_\mathrm{break}$ and
$P_\mathrm{central}$.

\subsection{Fitting the PLDF through Maximum Likelihood Estimation} \label{sec:MLE}

The PLDF has eight free parameters: $F_\mathrm{0}$, $P_\mathrm{break}$, $\beta_{1}$, $\beta_{2}$, $P_\mathrm{central}$, $s$, $\chi_1$, and $\chi_2$. To fit these parameters within each stellar mass bin, we adopt the maximum likelihood estimation (MLE) approach of \citet{Youdin2011} and \citet{Burke2015}. The Poisson likelihood function for a survey that detects $N_\mathrm{pl}$ planets around $N_*$ stars is given by Equation (18) of \citet{Youdin2011} and Equation (9) of \citet{Burke2015} as:
\begin{equation}\label{eqn:likelihood}
    L \propto \left[F_\mathrm{0}^{N_\mathrm{pl}}C_\mathrm{n}^{N_\mathrm{pl}}\prod_{i=a}^{N_\mathrm{pl}} g(\Porb,\Rp)\right] \exp(-N_\mathrm{exp}),
\end{equation}
where the PLDF predicts the number of survey detections with
\begin{equation}
    N_\mathrm{exp} = F_\mathrm{0} C_\mathrm{n} \int\limits_{P_{\mathrm{min}}}^{P_{\mathrm{max}}} \int\limits_{R_{\mathrm{min}}}^{R_{\mathrm{max}}} \left[ \sum_{j=1}^{N_*} \eta_j (\Porb,\Rp)\right] g(\Porb,\Rp) \mathrm{d} \Porb \mathrm{d} \Rp.
\end{equation}
In the above equation, the expected number of planets is converted to an expected number of detections by modifying the PLDF with the per-star survey effectiveness $\eta_j$ summed over $N_*$ sample targets (the term in brackets). Where \citet{Burke2015} treat $\eta_j$ as the pipeline completeness, we consider it as the detection efficiency and compute it separately for each stellar mass bin via \epos{}. 

We begin by specifying uniform priors for all parameters. By definition of $\chi_1$ and $\chi_2$ as fractions, their values are bound between 0 and 1. To reflect the behavior of the observed small planet population described in Section (\ref{sec:observed_turnover}), we also require that $\chi_1 > \chi_2$ such that super-Earths decrease in frequency around $P_\mathrm{central}$. We require some small orbital period region on either edge of our domain where $G$ is mostly flat - i.e. the smoothness around $P_\mathrm{central}$ is bounded to level off within a few days of either orbital period extreme. Because $P_\mathrm{central}$ is always a few factors of $s$ greater than $2$\,days and less than $100$\,days in $\log_{10}$ orbital period, we believe this is an appropriate requirement. We maximize the likelihood function (rather, minimize the negative log likelihood) through the Markov chain Monte Carlo (MCMC) program \emcee{} \citep{ForemanMackey2012}.

As mentioned in Section~(\ref{sec:model}), the likelihood function of Equation~(\ref{eqn:likelihood}) accounts for survey completness but not reliability. Therefore, we use the method of \citet{Bryson2020-OG-Reliability} outlined in Section~(\ref{sec:model}), where the model is fit repeatedly while varying the input planet sample as weighted linearly by their reliability scores. The posteriors of each fit are then concatenated to produce reliability-informed posteriors.

\begin{figure*}
    \centering
    \includegraphics[width = 0.8\textwidth]{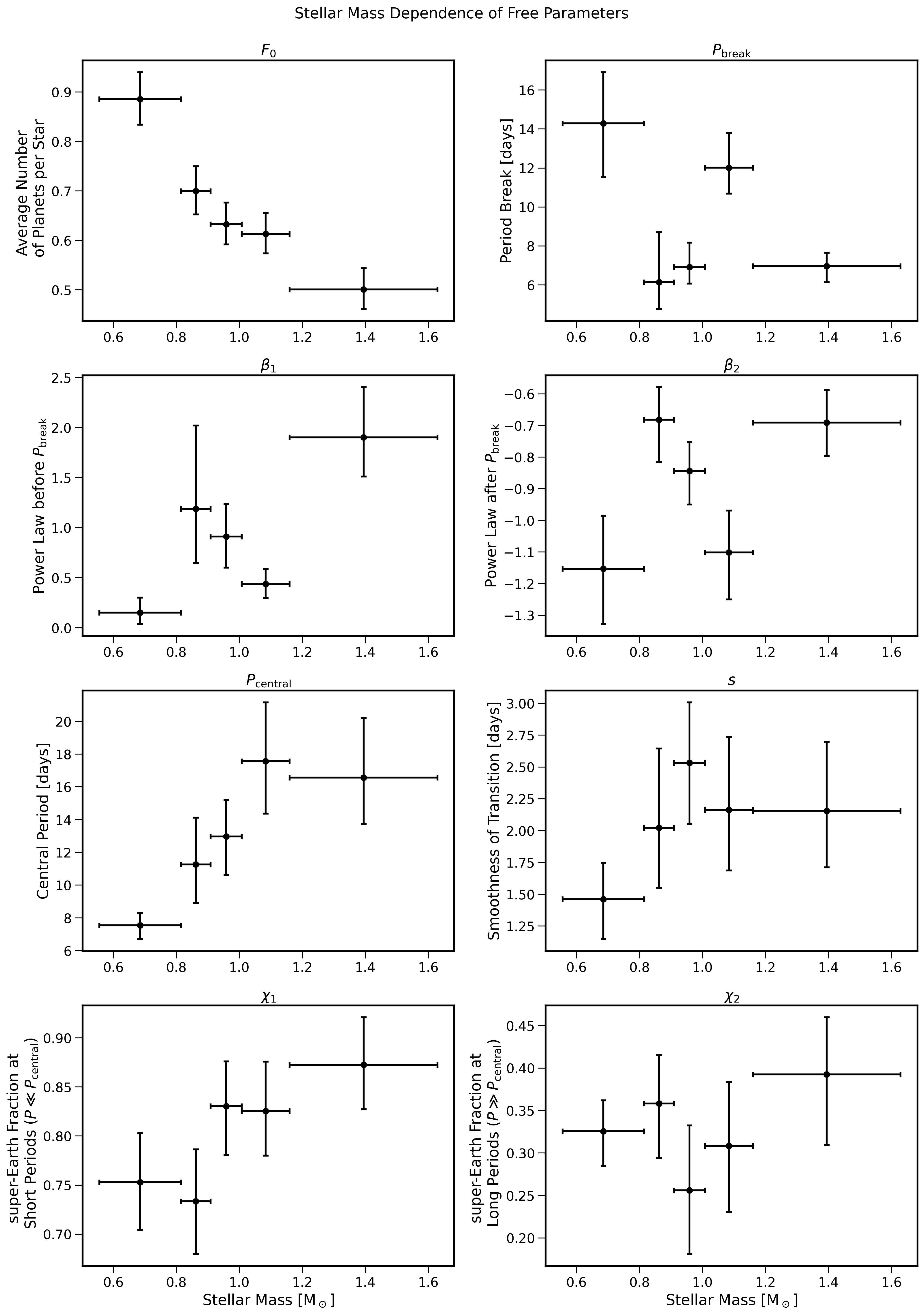}
    \caption{Best-fit parameters as a function of stellar mass. Each panel represents a different free parameter, and each mass bin was fit independently. Vertical error bars indicate $1\,\sigma$ uncertainties as the 16th and 84th percentile values from the posterior. Horizontal error bars represent the stellar mass range of each bin.}
    \label{fig:BestFit-vs-Mstar}
\end{figure*}

We take the median value of the \emcee{} posteriors for each parameter and report them in Table (\ref{tab:fits}) for each stellar mass bin. Using the MCMC approach allows us to numerically evaluate the posteriors and provide estimates of the $68\%$ credible intervals for each parameter. Some parameter posteriors are not conventional Gaussian distributions, so we use the phrase ``$1\sigma$" with caution. We include these uncertainties in Table (\ref{tab:fits}) and use them as vertical error bars in Figure (\ref{fig:BestFit-vs-Mstar}), which shows how all fit parameters vary across stellar mass bins. Aside from the stellar mass dependencies of $F_0$, $P_\mathrm{central}$ and $\chi_1$ listed previously, we note no clear stellar mass dependence beyond the typical uncertainties.

\begin{figure*}
    \centering
    \includegraphics[scale=0.4]{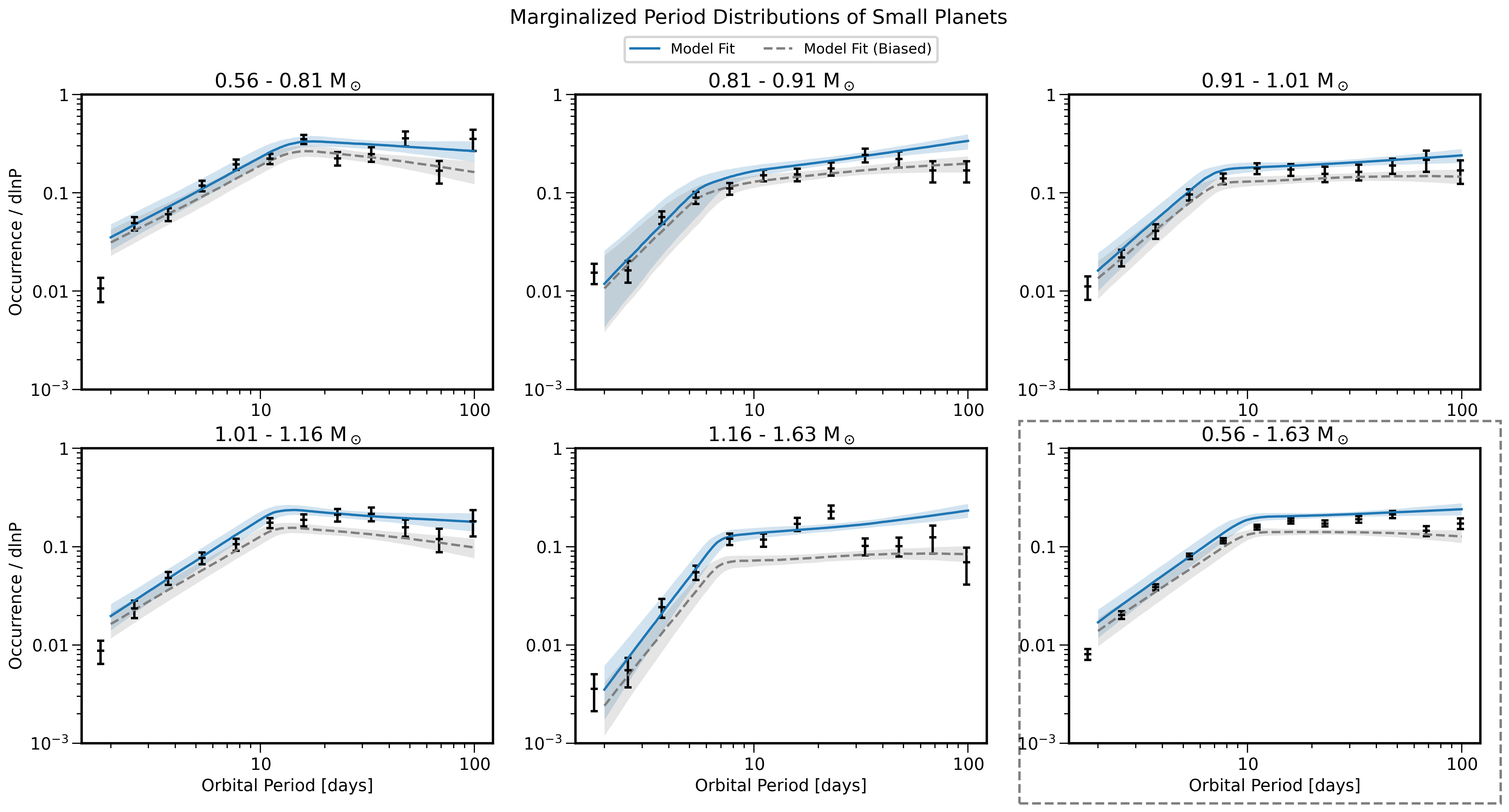}
    \caption{The occurrence of small planets marginalized over planet radius ($1 \leq \Rp \leq 3.5\,\EarthRad$) as a function of orbital period. Each panel represents a different stellar mass bin, with the full FGK sample shown in the lower-right. Black points are the observed marginalized occurrence as calculated by \epos{} for a selection of orbital period bins. Blue lines indicate the best fit distributions evaluated for that bin. Grey lines represent the best fit distributions scaled by the average detection efficiency computed for stars in that bin.}
    \label{fig:MarginalizedDist}
\end{figure*}

The marginalized orbital period distributions are calculated by using each bin's best-fit parameters to integrate Equation (\ref{eqn:df}) over the radius range in our sample. These distributions are presented in Figure (\ref{fig:MarginalizedDist}), where we include the observed marginalized occurrence rates for a number of orbital period bins calculated with \epos{}. The broken power law form chosen here is well-suited for describing the relationship between normalized occurrence and orbital period for the five stellar mass bins investigated here. We also include a ``biased" version of our fits, where the predicted occurrence is scaled by the detection efficiency, to illustrate that the inverse detection efficiency method typically underestimates occurrence in regions of low completeness.

The parameters governing the marginalized orbital period distribution are mostly consistent with previous works adopting the same form of a broken power law in orbital period. The power law exponents $\beta_1$ and $\beta_2$ (with an added $+1$ for the $\dd P$ to $\dd \ln{P}$ conversion) are similar to those of \citet{Mulders2018} (their Table~1) and \citet{Pascucci2019} (their Table~1). Both works fit directly to \epos{} occurrence rates while this work fits to \kepler data directly. In our largest mass bin, an unusually large value for $\beta_1$ produces a steeper slope which appears to match the data well. Values of $\beta_2$ in our first and fourth mass bins (lightest and super-Solar respectively) appear low, but are consistent with flat distributions to $1\sigma$ so we do not comment on implications of a decreasing distribution with orbital period. As for the orbital period break $P_\mathrm{break}$, many works suggest a value around $10$\,days as originally presented by \citet{Youdin2011} and \citet{Howard2012}. We find a dichotomy of orbital period breaks, with some at $\sim7$\,days and others at $\sim12$\,days, which appears independent of stellar mass. There is little evidence of a bimodal distribution in the posteriors of any one mass bin, which presents an interesting behavior meriting further study.

\section{Effects of Reliability Implementation}\label{ReliabilityEffects}

\begin{figure*}
    \centering
    \includegraphics[scale=0.4]{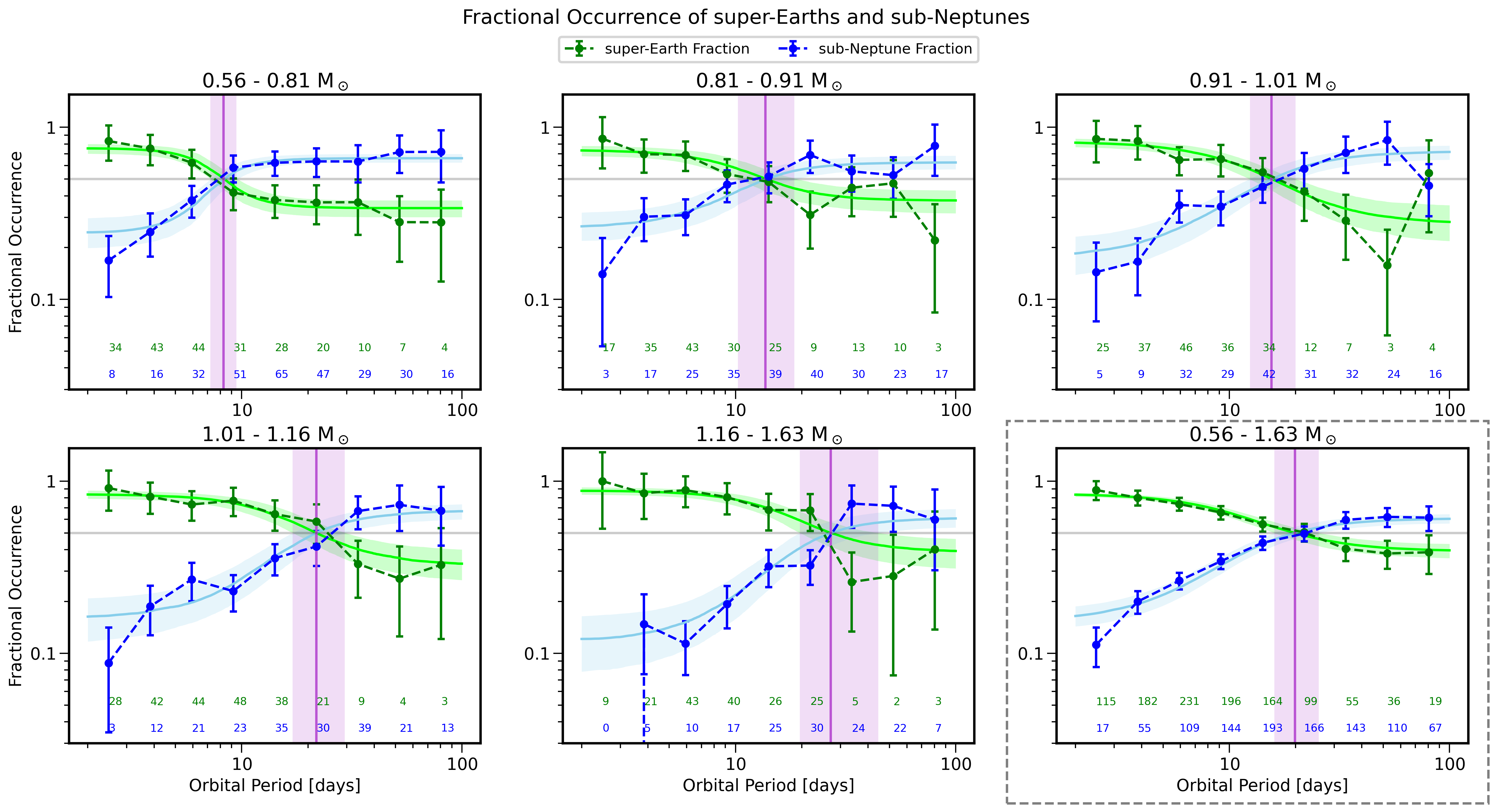}
    \caption{Similar to Figure~(\ref{fig:FracOcc}), but \textit{without} considerations of reliability for both the best-fit model and the \epos{}-derived occurrence rates. The results of the inverse detection efficiency method change more under the effects of reliability implementation than those of our forward model, likely due to the former's stronger susceptibility to biases in bins with a low number of observations.}
    \label{fig:noReliability}
\end{figure*}

To assess the impact of reliability on our forward model, we also perform a set of fits and inverse-detection efficiency calculations (via \epos{}) \textit{without reliability}. The latter follows the conventional inverse detection efficiency method, where the contribution of a planet candidate is not weighted by its reliability. For the former, likelihood estimation is performed in each stellar mass bin with $64$ walkers and $50,000$ steps (with $1,000$ discarded for burn-in). The results for both methods neglecting reliability are shown in Figure~(\ref{fig:noReliability}).

The model parameters are only slightly affected by reliability --- for example, incorporating reliability causes a $\sim6\%$ decrease in the value of $F_0$ for all stellar mass bins --- and all parameters remain consistent at the $1\sigma$ level to the case without reliability. This contrasts with the fractional occurrence rates derived from \epos{}, where the effects of reliability implementation may produce marked changes in a given bin - however, there is no clear trend in the direction of such changes regarding agreement with our forward model. This is partially due to our consideration of fractional occurrence: while incorporating reliability can only ever decrease a planet's contribution to the total occurrence, down-weighting a low-reliability sub-Neptune would increase the fractional occurrence of super-Earths (or vice versa) in a given bin of orbital period and stellar mass. Thus the effects of reliability implementation on the fractional occurrence rates computed with the standard method are dependent on the underlying reliability distribution of candidates, for which there is no discernible trend by virtue of reliability as a per-candidate metric.

The inverse detection efficiency method's susceptibility to reliability when compared to our forward model may be explained by noting that reliability is a statistical factor affecting the observed number of planets (directly used in the former), while our forward model attempts to constrain the intrinsic and unbiased population. It follows that reliability implementation should produce some of the strongest variations when considering the long-period population, for which there is a marked decrease in \kepler detection efficiency and thus a lower number of small planet detections. This is confirmed by a comparison with \citet{Bryson2020-OG-Reliability}, in which they fit to a further-out population ($50 < \Porb < 400$\,days) and find that $F_0$ decreases by $\approx30\%$ when implementing reliability, whereas we find only a $\sim6\%$ change in $F_0$ when considering the closer-in population within $100$\,days. Nevertheless, we follow the recommendations of previous occurrence rate studies and focus on the model(s) which include reliability in this work.

\end{document}